\documentclass{emulateapj-rtx4}
\pdfoutput=1

\newcommand{\mjyb}{mJy beam$^{-1}$}

\slugcomment{accepted in ApJ}

\shorttitle{Radio Relics in Galaxy clusters}
\shortauthors{Kale \& Dwarakanath}

\begin{document}

\title{Multi-frequency studies of radio relics in the galaxy clusters A4038, A1664 and A786}

\author{Ruta Kale$^{1,2}$ and K. S. Dwarakanath$^1$}

\affil{$^1$Raman Research Institute, C. V. Raman Avenue, Sadashivanagar, Bangalore 560 080}
\affil{$^2$Inter University Centre for Astronomy and Astrophysics, Post Bag 4, Ganeshkhind, Pune University Campus, Pune 411 007}

\email{ruta@iucaa.ernet.in}

\begin{abstract}
 We present a multi-frequency study of radio relics associated with the galaxy clusters A4038, 
A1664 and A786. Radio images, integrated spectra, spectral index maps and fits to the integrated spectra in the framework of the adiabatic compression model are presented. Images of the relic in A4038 at 150, 240 and 606 MHz with the GMRT have revealed extended ultra-steep spectrum ($\alpha \sim -1.8$ to $-2.7$) emission of an extent $210\times80$ kpc$^{2}$. The model of passively evolving radio lobes compressed by a shock best fits the integrated spectrum. The relic with a circular morphology at the outskirts of the cluster A1664 has an integrated spectral index of $\sim -1.10\pm0.06$ and is best fit by the model of radio lobes lurking for $\sim4\times10^7$ yr. The relic near A786 has a curved spectrum and is best fit by a model of radio lobes lurking for $\sim3\times10^7$ yr. At 4.7 GHz, a compact radio source, possibly the progenitor of the A786 relic, is detected near the center of the radio relic. The A786 radio relic is thus likely a lurking radio galaxy rather than a site of cosmological shock as has been considered in earlier studies. 
\end{abstract}


\section{Introduction} 

Diffuse extended radio sources with elongated or filamentary morphologies that cannot be identified with any unique optically detected galaxy are termed as radio relics (e.g. see the review by Ferrari et al. 2008). The arclike radio relics that are typically found around massive and dynamically disturbed galaxy clusters \citep[e.g.,][]{bag06, wee09} are proposed to be  electrons accelerated at shock fronts by the diffusive shock acceleration mechanism \citep{ens98}. The accelerated electrons could be those from the thermal pool of the galaxy cluster or the old mildly relativistic population due to active galaxies \citep{mar05}. Apart from the $\sim$ Mpc size arc-like relics, there are relics that have smaller sizes ($<500$ kpc) \citep[e.g.][]{sle01, wee09}. These have been proposed to be lurking lobes of radio galaxies or adiabatically compressed lobes of radio galaxies by shock waves \citep[]{ens01, ens02}. 

Regardless of their origin, radio relics are basically clouds of relativistic electrons and magnetic fields that evolve over timescales of $\sim$ ten million years which is the lifetime of the synchrotron emitting electrons. Synchrotron emission, adiabatic expansion and inverse Compton scattering of the CMB photons are the mechanisms by which the radio relics lose energy. The energy losses due to synchrotron and inverse-Compton scattering depend on the square of the energy of the electrons and thus produce steepening in their spectra. Therefore to study the evolution of the radio relics, a multi-frequency approach is essential. In a multi-frequency study by \citet[]{sle01}, integrated spectra of radio relics were modelled as evolving lobes of radio galaxies losing energy by synchrotron emission and inverse-Compton scattering (energy loss due to expansion was not considered). 

The spatial distribution of spectral indices over the extent of the relic gives important clues about its origin. For example, a relic created at a shock front shows an edge of young, flat spectrum plasma and a gradual steepening of spectra behind the edge. Such trends have been mapped in a few arclike double radio relics such as  A2345, A1240 \citep[]{bon09}, ZwCl $2341.1+0000$, CIZA J2242.8$+$5301, ZwCl $0008.8+5215$ \citep[]{wee09,wee10,wee11a} and a few arc-like single relics such as A2256 \citep[]{cla06,kal10}, A521 \citep[]{gia08} and A2744 \citep[]{orr07}.
 
Spectral index distribution studies of single radio relics are rare \citep[e.g.,][]{coh11,ran10, wee11b}. These relics show a wide variety in their morphologies and locations with respect to cluster centers. In this paper we present a multi-frequency study of three single radio relics which are associated with the galaxy clusters A4038, A1664 and A786. The relics and their host clusters are introduced in Sec. 1.1. The radio observations and data reduction are described in Sec. 2. Radio images of the relics are presented under Sec. 3. Integrated spectra are presented in Sec. 4 and model fits to the integrated spectra are described in Sec. 5. Results of each of the relic are discussed in Sec. 6 and a summary is presented in Sec. 7. We adopt a cosmology with $H_0=73$ km s$^{-1}$Mpc$^{-1}$, $\Omega_{matter}=0.27$ and $\Omega_{vacuum}=0.73$.

\subsection{The clusters and the relics}
The relic in A4038 (z$=0.03$ \citep[]{str99}, L$_{x[0.1-2.4keV~h_{50}^{-1}]} \sim 2.87\times10^{44}$ erg s$^{-1}$ \citep[]{rei02}) was discovered as an ultra-steep spectrum source in surveys (160 MHz) by \citet[]{sle83}. 
Based on further observations at 1.4 GHz \citep[]{sle84, sle98, sle01} (hereafter, SR98 and SL01), the extent of the relic was reported to be 56 kpc and a polarized intensity of $4\%$ was detected. Spectral index ($\alpha$, $S\propto\nu^{\alpha}$) over a narrow range of frequencies around 1.4 GHz was estimated to be -3.1.

In A1664 (z$=0.1241$ \citep[]{str99}, L$_x=5.36\times10^{44}$ erg s$^{-1}$ \citep[]{pim06}) a relic was identified from the NVSS by \citet[]{gio99}. It is  known for its circular morphology \citep[]{gov01} unlike other relics which have elongated morphologies. It has not been imaged at any frequencies other than 1400 MHz and thus the spectral index is unknown.
\par The relic near A786 was first identified and studied by \citet[]{dew91} and \citet[]{har93}. This relic is located at a distance of $\sim 5$ Mpc from the center of the cluster A786 (z$=0.128$  \citep[]{str99}, log $($L$_x) \sim 44.8$). The relic has been discussed as a site of a large scale cosmological shock \citep[]{ens98}. This relic is also referred to as `0917+75' in literature.
\section{Radio Observations and Data Reduction}
The Giant Meterwave Radio Telescope (GMRT) was used to carry out observations of the relics in the frequency bands of 150, 240, 325, 606 and 1288 MHz. The shortest baseline of 100 m and the longest baseline of 25 km of the GMRT in a hybrid configuration provide an opportunity to image extended sources such as radio relics with good resolution and sensitivity at multiple frequencies. The Westerbork Synthesis Radio Telescope (WSRT) was used to carry out observations of the relic in A786 at 345 MHz. Apart from these, archival data from the Very Large Array (VLA) at 4.7 GHz and images from the NRAO VLA Sky Survey (1.4 GHz, NVSS, \citet[]{con98}) and the VLA Low Frequency Sky Survey (74 MHz, VLSS, \citet[]{coh07}) were used.
A summary is given in Table 1. 

Reduction of all the data were carried out using NRAO AIPS (Astronomical Image Processing System). Standard steps of data editing and complex gain calibration were carried out. The data lost due to radio frequency interference (RFI) varied between $5 -35\%$ over the range of frequencies 1288 to 150 MHz at the GMRT. The WSRT observation of the relic in A786 consisted of 8 bands, each of band width 10 MHz, between 310 and 390 MHz. One frequency band of 10 MHz had to be discarded due to RFI. The latest VLA values were used for setting flux calibration (AIPS version 31DEC10). In the frequency bands, 150, 240, 325 and 606 MHz, wide field imaging was carried out to overcome the effects of non-coplanarity. The visibilities were weighted according to the 'ROBUST=0' scheme implemented in AIPS. Self-calibration was carried out to obtain deeper images at all the frequencies.

The angular extents of the relics ($\sim 5', 6'$ and $10'$, see Sec. 3) are much smaller as compared to the largest angular extents that can be fully imaged with the GMRT at 240 ($\sim 44'$), 325($\sim 32'$) and 606 ($\sim 17'$) MHz, with the WSRT at 345 ($\sim 95'$) MHz and with the VLA-D at 1400 ($\sim 15'$) MHz. Therefore no flux is missed due to missing short baselines. Images at multiple frequencies made with the same resolution, corrected for the primary beam gains and  blanked in regions with flux densities below $5\sigma$ were used to construct spectral index maps.
\section{Radio Images and Spectral Index Maps}
\subsection{Abell 4038}
Images of A4038 using the GMRT at 150 and 1288 MHz are presented in  Fig. 1 as contours and grey-scale respectively. The elongated, filamentary, `hook' shaped structure detected at 1288 MHz is the radio relic in A4038 as reported earlier in the literature (SR98). At 150 MHz extended emission about thrice the size of that at 1288 MHz is detected. The extent of the relic detected at 150 MHz is  133 kpc ($\sim 3.8'$) along N-S and $80.5$ kpc ($\sim2.3'$) along E-W. The positions of discrete radio sources (unresolved sources or those identified with optically detected galaxies) are marked by crosses. The northern cross marks a compact radio source which does not have any optical counterpart (A$4038\_{10}$, SL01) and is presumed to be a background source unrelated to A4038. The southern cross marks a radio source identified with the cD galaxy in A4038 (A$4038\_{11}$, SL01). In Fig. 2 (left), shown in grey-scale is the DSS R band image of the A4038 field. The cD galaxy can be clearly seen in the R band image. 
\par In the high resolution image at 240 MHz (Fig. 2, left) blobs of emission are seen scattered to the NW of the cD (A4038$\_11$) which indicate the presence of extended features. We produced low resolution ($25''\times25''$) images at 240 MHz (Fig. 2, right) to confirm the presence of the suspected extended features. We refer to the low surface brightness extension toward north west as the `NW-Extension' or `NW-Ext' for brevity, hereafter. The NW-Ext has an extent of 84$\times$56 kpc$^2$. The average surface brightness of the NW-Ext at 240 MHz is $\sim$7 mJy beam$^{-1}$ (beam$=25''\times25''$). The image at 606 MHz convolved to a resolution of $25''\times25''$ is presented in Fig. 3 (contours). The high resolution 606 MHz image is shown in grey-scale. The low surface brightness NW-Ext is not detected at 606 and 150 MHz. The surface brightness detections limited to $\sim2$ mJy beam$^{-1}$ at 606 MHz and to $\sim20$ mJy beam$^{-1}$ at 150 MHz resulted in non-detections of the NW-Ext at these frequencies. The non-detection of NW-Ext at 606 MHz implies an upperlimit on the spectral index of NW-Ext of  $\alpha_{240}^{606}<-1.3$. The NVSS \citep[]{con98} and the VLSS \citep[]{coh07} images at 1400 and 74 MHz respectively, are presented in Fig. 4. Extended emission is clearly detected in both the images.

We refer to the high surface brightness regions of the relic detected at the frequencies 150, 240 and 606 MHz (Figs. 1, 2 and 3) as the `main relic'. A reference to `the relic', hereafter will mean the `main relic' together with the NW-Ext. 

The spectral index map of A4038 between 606 and 240 MHz with a resolution of $25''\times25''$ and the corresponding uncertainty map are presented in Fig. 5. The positions of the discrete sources are marked by $'+'$s and are not subtracted from the images. The map shows a complex distribution of spectral indices. The central region of the relic has a spectral index of -1.3 and it steepens toward north and south reaching spectral indices of -2.7. 
\subsection{Abell 1664} 
Images of A1664 at 150 and 325 MHz using the GMRT and at 1400 MHz from the NVSS are presented in Fig. 6. The relic has a circular morphology and has an angular extent of $\sim 6'$ (linear extent $\sim0.927$ Mpc). To the north east, the relic appears connected to an unresolved radio source associated with the galaxy A1664$\_3$ \citep[]{pim06}. The high resolution image at 325 MHz with the GMRT resolves this compact source (Fig. 7).  
The spectral index map between 325 and 1400 MHz and the corresponding uncertainty map are given in Fig. 8. The relic has a flat spectral index distribution ($\sim-0.5$ to $-1.0$) with small pockets of steep spectral index. The source A1664$\_3$ has a spectral index of -0.5.
\subsection{Abell 786}
The images of the relic associated with A786 at 150, 345, 606 and 1400 MHz are presented in Fig. 9. The largest extent of the relic is detected at 345 MHz ($\sim10'$). A central region with higher surface brightness is seen at 345 and 1400 MHz with low surface brightness extensions along east and west. A large fraction ($\sim40\%$) of short baseline data at 150 MHz was corrupted due to RFI and had to be discarded. The sensitivity in the 150 MHz image is $\sim10$ \mjyb. The bright central portion of the relic is detected but the weak extensions along the east and west are not detected  at 150 MHz. To the southwest of the relic is a bright unresolved radio source (source R5, \citet[]{har93}). It is presumably a background source not related to the relic.
 
A high resolution image at 606 MHz ($7''\times4''$, PA = 0.16$^\circ$) is shown in Fig. 10 (left). Two unresolved sources (marked as 'twin sources' in Fig. 10 and by $'+'$ in Fig. 9) were detected at the northern edge of the relic. The positions (J2000) of these sources are RA 09h22m19.096s DEC +75d01$'$41.95$''$ and RA 09h22m24.77s DEC +75d01$'$43.92$''$ (separation $\sim20''$). No counterparts to these are found in the DSS and 2MASS images. We also produced an image at 4.7 GHz using archival data from the VLA (D-array) (Fig. 10, right). A compact source is detected in the central region of the relic (RA  09h22m09.27s DEC 74d$59'48.99''$, J2000) with extended emission of linear extent $\sim 350$ kpc surrounding it. A counterpart to this source is detected in 2MASS (total magnitudes in J, H and K$_s$ bands are 13.976, 13.584 and 12.701, respectively), however no information on the redshift of it is available. From the K$_s-$z correlation \citep[]{deb02}, the K$_s$ band magnitude implies a redshift of $\sim 0.1$ (redshift of A786 is 0.128).
Spectral index maps between 345 and 606 MHz and between 606 and 1400 MHz were produced and are shown in Figs. 11 and 12 respectively. The distribution of spectral indices over the relic is complex.
\section{Integrated Spectra}
\subsection{A4038}
 The total flux density in a box enclosing the relic and two discrete sources (A$4038\_11$ and A$4038\_10$) in A4038 was found using the images with resolutions $25''\times25''$ at each of the frequencies (150, 240 and 606 MHz). Total flux densities were also estimated at 1400 and 74 MHz using the NVSS  and the VLSS images, respectively, using boxes of same dimensions. 
To obtain the flux density of only the relic, contribution from the sources A$4038\_11$ and A$4038\_10$ had to be subtracted from the total flux density in the box. 
 The high resolution images at 1288 and 606 MHz were used to find the flux densities of the sources A$4038\_11$ and A$4038\_10$. Since these sources are embedded in the diffuse emission of the relic at 74, 150, 240 and 1400 MHz and could not be resolved, the spectral index between 1288 and 606 MHz was used to estimate their contributions at the other frequencies.
 The spectral indices ($\alpha^{1288}_{606}$) of A$4038\_11$ and A$4038\_10$ were found to be -0.55 and -0.88, respectively. It was assumed that the spectral indices are constant over the frequency range of 74 to 1400 MHz. 
 The flux densities at 29, 327, 408 and 843 MHz corrected for the contributions from the sources A$4038\_11$ and A$4038\_10$ and also other sources in the beams as reported by SL01 are given in Table 2. The integrated spectrum of the relic is plotted in Fig. 13 (points). The power of the relic at 1.4 GHz is $1.13\times10^{23}$ W Hz$^{-1}$.

The flux densities of the relic at 80 and 160 MHz are reported to be $19.0\pm2.7$ Jy and $4.3\pm0.5$ Jy respectively, by SL01. However since we have better estimates at 74 and 150 MHz (Table 2), we do not use the SL01 values at 80 and 160 MHz. 
\subsection{A1664}
The flux density in a box enclosing the relic and the radio source A1664$\_3$ was estimated. From higher resolution images, the flux density of A1664$\_3$ was estimated and subtracted. These flux densities are listed in Table 3. The integrated spectrum of the relic is plotted in Fig. 15 (points).  The spectral indices of the relic, $\alpha_{150}^{325}$ and  $\alpha_{325}^{1400}$ are $-1.32\pm0.03$ and $-0.98\pm0.01$, respectively. The power of the relic at 1.4 GHz is  $3.41\times10^{24}$ W Hz$^{-1}$. 
In the highest resolution map ($12''\times6''$, rms 0.6 \mjyb) at 325 MHz, an unresolved  source is detected in the region of the relic (marked `$+$' in Fig. 6). The position (J2000) of the source is RA 13h03m30.379s DEC $-24$d$21'54.29''$ and the peak and the integrated flux densities are $6.14\pm0.06$ and $8.1\pm1.3$ \mjyb, respectively (about $2\%$ of the total flux density of the relic). This source does not have a detectable counterpart in the DSS R band image and in the 2MASS J, H and K$_s$ band images. 
\subsection{A786}
The flux density in a box enclosing the relic, the twin discrete sources and the central source detected at 4.7 GHz was determined from the images at 150, 345, 606 and 1400 MHz. 
The total flux densities of the twin discrete sources in the relic region detected at 606 MHz and 4.7 GHz are 7.6 mJy and 1.5 mJy, respectively. Using the spectral index of -0.78 between 606 MHz and 4.7 GHz of the twin sources, the flux densities at 150, 345 and 1400 MHz were estimated and subtracted from the flux densities of the relic. The integrated flux densities of the relic after this correction are reported in Table 3. The power of the relic at 1.4 GHz is $3.59\times10^{24}$ W Hz$^{-1}$. The source detected at 4.7 GHz in the central region of the relic has a peak flux density of $0.34\pm0.06$ \mjyb and a total (including the surrounding extended emission) flux density of $7.2\pm0.3$ \mjyb. Since the full extent of the relic is not detected at 4.7 GHz we do not report total flux density of the relic at 4.7 GHz. The central compact source is not detected at lower frequencies, therefore no correction due to this source is made to the flux densities of the relic reported at lower frequencies. The integrated spectrum of the relic is plotted in Fig. 16 (points). 
\section{Adiabatic compression model fits}
Radio relics are proposed to be either remnants of radio galaxies or sites of accretion/ merger shocks where particles are accelerated to relativistic energies. The remnants of radio galaxies can be ageing lobes or lobes compressed by shocks in the intracluster medium (ICM). In the case of A4038, SL01 attempted a fit considering the relic as a remnant of a radio galaxy. The SL01 fits were based on estimates of volume and magnetic field using the extent of the relic (LLS$=$56 kpc) as detected by SL01 at 1.4 GHz. This extent of the relic is an underestimate due to the discoveries of extended features of the relic at lower frequencies. Further the models considered by SL01 neglected energy losses due to expansion of the lobes. We describe our attempt to fit the integrated spectrum of the relic with a model which overcomes the above mentioned drawbacks.
 
Revival of fossil radio lobes by adiabatic compression by passage of shocks is one of the models favored for radio relics \citep[]{ens01} (EG01 hereafter). The framework of EG01 considers revival of fossil radio cocoon by adiabatic compression due to the passage of shocks in the ICM.
The evolution of the spectrum of a radio galaxy before and after the jets cease to be active and after adiabatic compression of the lobes can be obtained. Energy losses due to synchrotron, inverse Compton (IC) and expansion are accounted for. There are five phases of evolution, viz. injection, expansion, lurking, flashing and fading. The salient features of the model and the application in the context of radio relics can be found in \citet[]{kal09}. 
 
The largest linear extent of the relic in A4038 is 210 kpc.
It is is located close to the cD galaxy and resides almost at the cluster center (in projection). 
 The volume of the complete relic was approximated to be the sum of the volumes of the main relic ($110\times75\times75$ kpc$^3$) and that of the NW-Ext (25$^3$ kpc$^3$).
 A magnetic field of $3\mu$G was estimated using the minimum energy condition under standard assumptions. 

Best fits to the integrated spectrum of A4038 relic in `flashing' and `fading' phases were obtained. The reduced Chi-square ($\chi^2_{red}$) was estimated to be 5.96 and 5.49 in the two fits respectively. The parameters used in the model for the best fit are listed in Table 4. The model fits along with the observed spectrum are plotted in Fig. 13 and 14 (left). The best fits are denoted by black lines. The model spectra in other phases are also plotted in the same figures for comparison. The right hand side panels in Figs. 13 and 14 show the best fits along with model spectra obtained when the values of the duration of the phases were changed by  $\pm30\%$. Note the deviation from the observed spectrum.
 
The relic in A1664 is at a distance of $\sim1.13$ Mpc from the center of the cluster. 
Best fit to the spectrum was found when the relic was assumed to be in the `lurking' phase (Fig. 15). The best fit parameters are listed in Table 4. The duration of the `lurking' phase was found to be $4\times10^7$ yr.
 
The EG01 model was used to fit the integrated spectrum of the relic near A786. A best fit was obtained when the relic was assumed to be in the `lurking' phase. The best fit parameters are reported in Table 4 and the fit is shown in Fig. 16 (left). The plots showing deviation from the best fit when the timescale was changed by $\pm 30\%$ are shown in Fig. 16 (right). The duration of the lurking phase phase was found to be $6\times10^7$ yr.
\section{Discussion}
The imaging of the three radio relics in galaxy clusters A4038, A1664 and A786 at low frequencies ($< 1$ GHz) has revealed several new properties of the relics. Images of comparable sensitivity and resolution were produced by the use of the GMRT, the WSRT and the VLA over the frequency range of 150-1400 MHz. The integrated spectra sampled at more than three frequencies were produced. The spectra are curved and steep ($\alpha<-1$) and unlike a flat power-law expected of young freshly accelerated plasma ($\alpha\sim-0.7$). The adiabatic compression model was used to fit the integrated spectra. The models of radio galaxy cocoons of ages $\gtrsim10^7$ yr fit the spectra. The spectral index maps show complex distributions. The gradients in the spectral index maps do not show edges of flat spectral indices that are characteristic of plasma accelerated at shock fronts. The spectral indices are steep and vary across the extent of the relic by as much as $1$ ($\Delta\alpha\sim1$). These properties imply that these relics contain aged relativistic plasma and can be termed as `relics of radio galaxies'. The model of shock accelerated plasma is disfavored by the observations. The properties of the relics are discussed below.
\subsection{Abell 4038}
Our sensitive low frequency observations with the GMRT have led to the discovery of steep spectrum emission in addition to the relic known from earlier high frequency studies. The largest linear extent of the relic in A4038 as seen at 240 MHz is $\sim210$ kpc as opposed to 56 kpc detected at 1.4 GHz. The relic is elongated along the N-S direction. A low surface brightness extension to the relic toward the NW has been discovered at 240 MHz. Due to limitations of the sensitivities of observations at 150 and 606 MHz, the NW-Ext was not detected at these frequencies. Deeper observations at these frequencies are required to detect and determine the spectral index of the NW-Ext.

We used the adiabatic compression model (EG01) to fit the integrated spectrum of the relic.
The EG01 model accounts for the energy losses due to expansion and also the resulting change in the magnetic field due to change in volume. Further, it also considers the case in which a shock in the ICM compresses lurking radio cocoons and leads to enhanced emission from them. The best fits were obtained in the `flashing' and the `fading' phases.
The durations of the phases of injection, expansion and lurking are the same in both the cases. However the extent of the compression in the `flashing' phases differs. In the case of `flashing' phase being the best fit, the volume changes by a factor of $\sim9.3$, whereas in the case of `fading' being the best fit, the change is by a factor of $\sim3.5$. The relation between the compression factor ($C$) and the pressure ratio ($P_1/P_2$) is given by, $C=(P_1/P_2)^{3/4}$. This implies that the pressure jumps at the shock that caused the compression are factors of $\sim20$ and $\sim 5$ in the `flashing' and the `fading' best fit cases, respectively.  
Using the pressure ratios and the factor $\gamma=5/3$ (adiabatic index for thermal gas), the Mach number $M$ (Eqn. 22, EG01) was found to be $\sim4$ and $\sim2$ in the cases of `flashing' and `fading' best fit respectively. 
It should be noted that the role of shock in this context is to compress the cocoon. The speed of sound in the radio cocoon being very high($\sim c/\sqrt{3}$), the relativistic plasma is not shocked, but is compressed as a result of an encounter with a shock (EG01). 
No direct acceleration at the shock front is believed to be operative.
The change in the spectral luminosity by factors of 10 is achieved in the compression phase making the relic detectable over the observed frequency range. 

The remnants of radio galaxies in clusters have been found in the form of bubbles in the ICM (eg. \citet[]{mcn02}). The jets of radio galaxies inflate the bubbles in the ICM which are detected in high resolution X-ray images as cavities. The cavities (bubbles) are often found to be filled with relativistic plasma detectable at radio wavelengths. These bubbles can get detached from the place where they formed and rise in the ICM due to buoyancy. The main relic in A4038 could be such a bubble and the NW-Ext could be an older bubble that has risen by buoyancy. In a study of radio bubbles in galaxy clusters, A4038 has been classified as a source with a central radio source but no detected cavities in X-ray observations with Chandra \citep{dun05}. In a later study it has been remarked that the sensitivity of the X-ray observations may have been insufficient in revealing the cavities since no extended radio emission was present to provide clues about the location of cavities \citep{dun06}. Given that we have detected extended steep spectrum radio emission in A4038, a re-analysis of the existing X-ray observations or a deeper observation may lead to detection of cavities.

The relativistic plasma that forms the relics has been considered to originate from a radio galaxy in the EG01 model. Over time, the host galaxy would have moved and may not have an AGN any more, preventing identification. The radio images of A4038 relic indicate a diffuse bridge connecting the relic to the cD. The bridge cannot be an active jet considering that it has a steep spectral index and is not detectable at higher frequencies (young relativistic plasma in the jet would have a flat spectrum). The cD galaxy could be the progenitor of the relic but the evidence is not sufficient to rule out other progenitors.
\subsection{Abell 1664}
The 150 and 325 MHz images presented above are the first low frequency images of the relic in A1664. These have yielded an estimate of the spectral index of the relic between 150 and 1400 MHz $\sim-1.10\pm0.06$. We attempted a model fit to the spectrum using the EG01 model. The best fit found in the `lurking' phase implies that no shock compression is required to produce the relic. This is consistent with the fact that there has not been any recent dynamical activity in the cluster as also implied by the cooling flow at the center of A1664 \citep[]{all95}. The cluster A85 which is also a cooling flow cluster hosts a peripheral relic. The main difference between A85 and A1664 relics is that the relic in A85 has an extent of $\sim 100$ kpc whereas A1664 relic has an extent $\geq0.92$ Mpc. 

The NVSS image overlaid on ROSAT (0.1-2.4 keV) image of A1664 is shown in Fig. 17. The cluster center is to the north of the relic. 
The X-ray surface brightness in the cluster center (Chandra image) is elongated and can be approximated to be an ellipse, the major axis of which is at a P. A. of 25$^\circ$ \citep[]{kir09}. The relic is situated along the same axis toward SW at a distance of 1.13 Mpc from the cluster center. The elongation in X-rays is consistent with the optical major axis of the galaxy at the cluster center and thus the elongation may be due to processes local to the cD or a non relaxed state \citep{pie98}. 
The location of the relic at the outskirts of the cluster and close to a radio source is similar to the relic in Coma cluster. The Coma cluster relic is at a distance of 2.2 Mpc from the cluster center and is possibly associated with a radio galaxy seen close to the relic (Ensslin et al. 1998; EG01). The cosmological structure formation shocks are believed to be found at cluster peripheries. These could compress cocoons of radio galaxies leading to the formation of relics and/or could accelerate the  particles in the thermal gas to form relics. The Coma relic is elongated and has been proposed to have been compressed by an accretion shock \citep[]{ens98}. The A1664 relic  shows similarity in terms of its location relative to the cluster center to the Coma relic, but it is morphologically different and may have a completely different origin.

The relic appears connected to the galaxy A1664$\_3$ in the images. However, no evidence for jets from the galaxy connecting to the relic has been found. It is possible that the A1664 relic is a blob of radio emission from the past activity of an AGN in the cluster. 
Deeper X-ray studies in the region of the relic are required to detect if there is a cavity corresponding to the A1664 relic. The model fit to the spectrum in the EG01 framework and the morphology of the A1664 imply that the relic is a lurking lobe of a radio galaxy.
\subsection{Abell 786}
The relic is about 5 Mpc from the center of the cluster A786 in the plane of the sky. In Fig. 18 the 345 MHz contours are overlaid on the ROSAT HRI image (0.1-2.4 keV). There are two optical galaxies near the brightest central region of the relic, called D and E by \citet[]{har93}, which have redshifts consistent with that of the cluster A786. For this reason the relic has been considered associated with the cluster A786. 

The integrated spectrum of the relic near A786 is curved. The spectral index between 150 and 606 MHz is $\sim-0.80$ and it steepens to a value of $\sim-1.27$ between 606 and 1400 MHz. A best fit in `lurking' phase was obtained. According to the model fit the radio galaxy responsible for depositing the relativistic plasma in the ICM  had been active for $\sim1.5\times10^7$ yr. The total time elapsed since the jets of the radio galaxy ceased to be active is $\sim5\times10^7$ yr. Within this timescale the progenitor galaxy would have travelled a distance of 50 kpc with a velocity of 1000 km s$^{-1}$ (typical at cluster center) and 5 kpc with a velocity of 100 km s$^{-1}$ (typical velocities in galaxy groups). Since at a distance of 5 Mpc from the cluster center the velocities of galaxies are low (similar to galaxy groups), it is likely that the progenitor lies within the extent of the relic. The compact radio source with extended features surrounding it, detected at 4.7 GHz could be the progenitor AGN and the relic, a fossil lobe associated with it.
 
The relic near A786 is one of the largest diffuse emission (LLS$\sim1.6$ Mpc) regions known in the Universe. Such large scale filamentary radio emission is rare and has been associated with filaments of galaxies or cluster peripheries (eg. ZwCl 2341.1+0000, Bagchi et al. 2002 and A3376, Bagchi et al. 2006). Structure formation shocks are believed to be responsible for such sources. \citet[]{ens98} have discussed this relic as a candidate for accretion shock compressed relic. However due to lack of information regarding the temperature and velocity dispersion of the cluster A786, the expected shock radius and polarization could not be determined \citep[]{ens98}. The spectral index maps of the relic presented here do not show gradients as would be expected in the case of shock acceleration. Also the integrated spectrum of the relic is consistent with that of a `lurking' radio galaxy lobe. Thus, it seems that this relic is not a site of cosmological structure formation shock acceleration. 
\section{Summary}
A multi-frequency study of the three radio relics associated with the galaxy clusters, A4038, A1664 and A786 is presented. Radio images and spectral index maps constructed from those are presented and discussed. We have used the adiabatic compression model to fit the integrated spectra of the relics. 

The 150, 240 and 606 MHz images of A4038 have revolutionized the picture of the relic known earlier. New features of the relic are uncovered extending the largest detected linear size of the relic from 56 kpc to 210 kpc. The newly discovered emission has spectral indices $\alpha\leq-1.8$.
The model of radio lobes compressed by shock best fits the integrated spectrum of A4038 relic.
The relic in A1664 was imaged at 150 and 325 MHz with the GMRT. These images and the NVSS image at 1400 MHz were used to construct the integrated spectrum of the relic in A1664.
A best fit to the integrated spectrum of the relic in A1664 was obtained in the `lurking' phase.
The images of the relic near A786 at 150, 345, 606 and 1400 MHz were used to construct the integrated spectrum. The integrated spectrum was best fit in the EG01 framework in the `lurking' phase. A source detected at 4.7 GHz in the central region of the relic that has an optically identified galaxy as a counterpart may be the progenitor of this relic. The spectral properties of the relic and the detection of a possible progenitor at the center of the relic imply that the relic is an old radio galaxy rather than a site of cosmological shock as considered in earlier studies.

\acknowledgments

We thank M. Johnston-Hollitt for suggesting the use of 4.7 GHz data for the A786 relic. We thank the anonymous referee for useful remarks and comments. We thank the staff of the GMRT who have made these observations possible. GMRT is run by the National Centre for Radio Astrophysics of the Tata Institute of Fundamental Research. The National Radio Astronomy Observatory is a facility of the National Science Foundation operated under cooperative agreement by Associated Universities, Inc. The Westerbork Synthesis Radio Telescope is operated by the ASTRON (Netherlands Institute for Radio Astronomy) with support from the Netherlands Foundation for Scientific Research (NWO). This research has made use of the NASA/IPAC Extragalactic Database (NED) which is operated by the Jet Propulsion Laboratory, California Institute of Technology, under contract with the National Aeronautics and Space Administration. This publication makes use of data products from the Two Micron All Sky Survey, which is a joint project of the University of Massachusetts and the Infrared Processing and Analysis Center/California Institute of Technology, funded by the National Aeronautics and Space Administration and the National Science Foundation. The Digitized Sky Surveys were produced at the Space Telescope Science Institute under U.S. Government grant NAG W-2166. The images of these surveys are based on photographic data obtained using the Oschin Schmidt Telescope on Palomar Mountain and the UK Schmidt Telescope. The plates were processed into the present compressed digital form with the permission of these institutions. This research has made use of data and/or software provided by the High Energy Astrophysics Science Archive Research Center (HEASARC), which is a service of the Astrophysics Science Division at NASA/GSFC and the High Energy Astrophysics Division of the Smithsonian Astrophysical Observatory.

\begin{table}
\begin{center}
\caption{Summary of Observations.\label{tbl-1}}
\begin{tabular}{ccccccc}
\hline 
Source & Freq., BW  & Telescope & Time & Synthesized & P.A.& rms  \\ 
       &  (MHz)     &           & (min) & beam($''\times''$) &($^\circ$) & 
       (mJy beam$^{-1}$) \\ 
       \hline
       & 74              & VLA-B (VLSS)  & -         &$80\times80$&0 &100\\  
A4038  & 150 , 6            & GMRT & 40        & $25\times25$&0 & 4.0 \\ 
       & 240 , 8            & GMRT & 240       & $15\times10$ &-11& 1.0\\ 
       & 606 , 16            & GMRT & 240       & $6\times3$ &28 &0.26\\ 
       & 1288 , 32           & GMRT & 200       & $3\times2$ &35 &0.093\\ 
       & 1400            & VLA-D (NVSS)  & -         & $45\times45$& 0 & 0.45\\ 
       \hline 
A1664  & 150 , 6            & GMRT         &210 &$53\times45$ &4 &7.0\\ 
       & 325 , 16            & GMRT         &120&$ 39\times30$&40 &1.0\\ 
       & 1400            & VLA-D (NVSS) & -      & $45\times45$ & 0 &0.3\\ 
 \hline
A786 & 150 , 6 & GMRT & 210 &$67\times52$&22&10.0\\  
     & 345 ,16& WSRT & 544 & $52\times51$ &7&0.8\\ 
     & 606 ,16& GMRT & 240 & $64\times64$ &0&2.3\\ 
     & 1400 & VLA-D(NVSS) & - &$45\times45$ &0&0.45 \\
     & 4700, 50 & VLA-D(AJ283) & 400 & $11\times11$ & 0&0.015\\
 \hline
\end{tabular}
\end{center}
\end{table}

\begin{table}
\begin{center}
\caption{Integrated flux densities of the relic in A4038.\label{tbl-2}}
\begin{tabular}{ccccc}
\hline 
Freq. (MHz)& Flux density  & A$4038\_11$        & A$4038\_10$         &relic\\
           & (total) (Jy)    & (mJy)                & (mJy)                 & (Jy)\\  
\hline
29         &$42\pm7$       &$- $                  &$- $                 & $32\pm7$\tablenotemark{a}\\
74	   &$12.6\pm1.5$   &$117\pm20$      &$24\pm5 $    &$12.45\pm1.5$\tablenotemark{b}\\
150	   &$5.26\pm0.11$  &$84\pm10$	&$13\pm2$      &$5.16\pm0.11 $\\
240        &$3.04\pm0.06$  &$65\pm5$     &$9.0\pm2.0$      &$2.96\pm0.06$\\
327        &$1.54\pm0.15$  &$-$          &$-$&    $1.44\pm0.15$\tablenotemark{a}\\
408        &$0.96\pm0.11$  &$ -$&$-$ & $0.91\pm0.11$\tablenotemark{a}\\
606        &$0.427\pm0.008$&$39.0\pm3.0 $    &$4.5\pm0.8 $   &$0.380\pm0.057$\\
843        &$0.21\pm0.03$  & $-$                 &$-$      &$0.17\pm0.03$\tablenotemark{a}\\
1288       &-&$25.0\pm2.0 $&$2.1\pm0.5$    &$-$\\
1400       &$0.086\pm0.004$&$23.0\pm0.1$ &$2.3\pm0.1$&$0.060\pm0.004$\tablenotemark{c}\\
 \hline
\end{tabular}
\end{center} 
\tablenotetext{a}{Table 3 in \citet[]{sle01}}
\tablenotetext{b}{VLSS}
\tablenotetext{c}{NVSS}
\end{table}

\begin{table}
\begin{center}
\caption{Integrated flux densities of the relics in A1664 and in A786.\label{tbl-3}}
\begin{tabular}{ccccc}
\hline 
Freq. (MHz)& A1664$\_3+$relic  & A$1664\_3$        & A1664 relic         &A786 relic\\
           & (total) (Jy)    & (mJy)                & (mJy)                 & (mJy)\\  
\hline
150	   &$1.811\pm0.181$ &$561.0\pm56.0$ &$1250.0\pm125.0$      &$933.5\pm90.0$\\
325        &$0.688\pm0.030$  &$238.0\pm10.0$ &$450.0\pm30.0$&- \\
345        &-              &-             & -             &$440.2\pm30.0$\\
606        &-&- &-    &$306.0\pm8.0$\\
1400       &$0.178\pm0.004$&$71.8\pm1.0$&$106.2\pm4.0$&$105.0\pm5.0$\\	
 \hline
\end{tabular}
\end{center} 
\end{table}
\clearpage
\begin{table}
\centering
\caption{Adiabatic compression model fit parameters for the relics.\label{tbl-4}}
\begin{tabular}{llllllll}
\hline
&Phase&          &$\Delta t$  &$\tau $ &$b$  & $V$       & $B$ \\
&     &     &(Gyr)         &(Gyr)     &     &(Mpc$^{3}$)& ($\mu$G)\\ \hline     
A4038&0&Injection   &0           &0.005   &1.8     &  0.0033       & 0.95 \\ 
&1&Expansion  &0.0054      &0.01    &1.2     &  0.0056       &  0.67    \\ 
&2&Lurking   &0.05         &$\infty$  & 0    & 0.0056        & 0.67 \\
&3&{\bf Flashing}   &0.37       &-0.55   & 2.0    &   0.0006      & 3.00 \\
&4&Fading  &0.1        &$\infty$     &0     &  0.0006     & 3.00\\
\hline
\hline
A4038&0&Injection   &0           &0.005   &1.8     & 0.0012   & 1.88 \\ 
&1&Expansion  &0.0054      &0.01    &1.2     & 0.0021    & 1.33    \\ 
&2&Lurking   &0.05         &$\infty$  & 0    &  0.0021   & 1.33 \\
&3&Flashing   &0.1       &-0.22   & 2.0    &  0.0006     & 3.00 \\
&4&{\bf Fading}  &0.067        &$\infty$     &0     &  0.0006 &3.00\\
\hline
\hline
A1664&0&Injection   &0           &0.015   &1.8    &  0.104     &   0.96 \\ 
&1&Expansion  &0.005      &0.04    &1.2      & 0.120     &   0.88  \\ 
&2&{\bf Lurking}   &0.04         &$\infty$  & 0& 0.120   &  0.88 \\
&3&Flashing        &0.03       &-0.17 & 2.0 &  0.081&  1.14\\
&4&Fading  &0.3        &$\infty$     &0      &  0.081    & 1.14\\
\hline
\hline 
A786&0&Injection   &0           &0.015   &1.8    & 0.223 & 1.87  \\ 
&1&Expansion  &0.012      &0.01    &1.2      &   0.576    & 1.00 \\ 
&2&{\bf Lurking}   &0.028         &$\infty$  & 0&   0.576  &1.00 \\
&3&Flashing        &0.02       &-0.04 & 2.0 & 0.144     & 2.51\\
&4&Fading  &0.06        &$\infty$     &0      &0.144     &2.51 \\
\hline
\end{tabular}
\end{table}

\begin{figure}
 \epsscale{0.5}
 \plotone{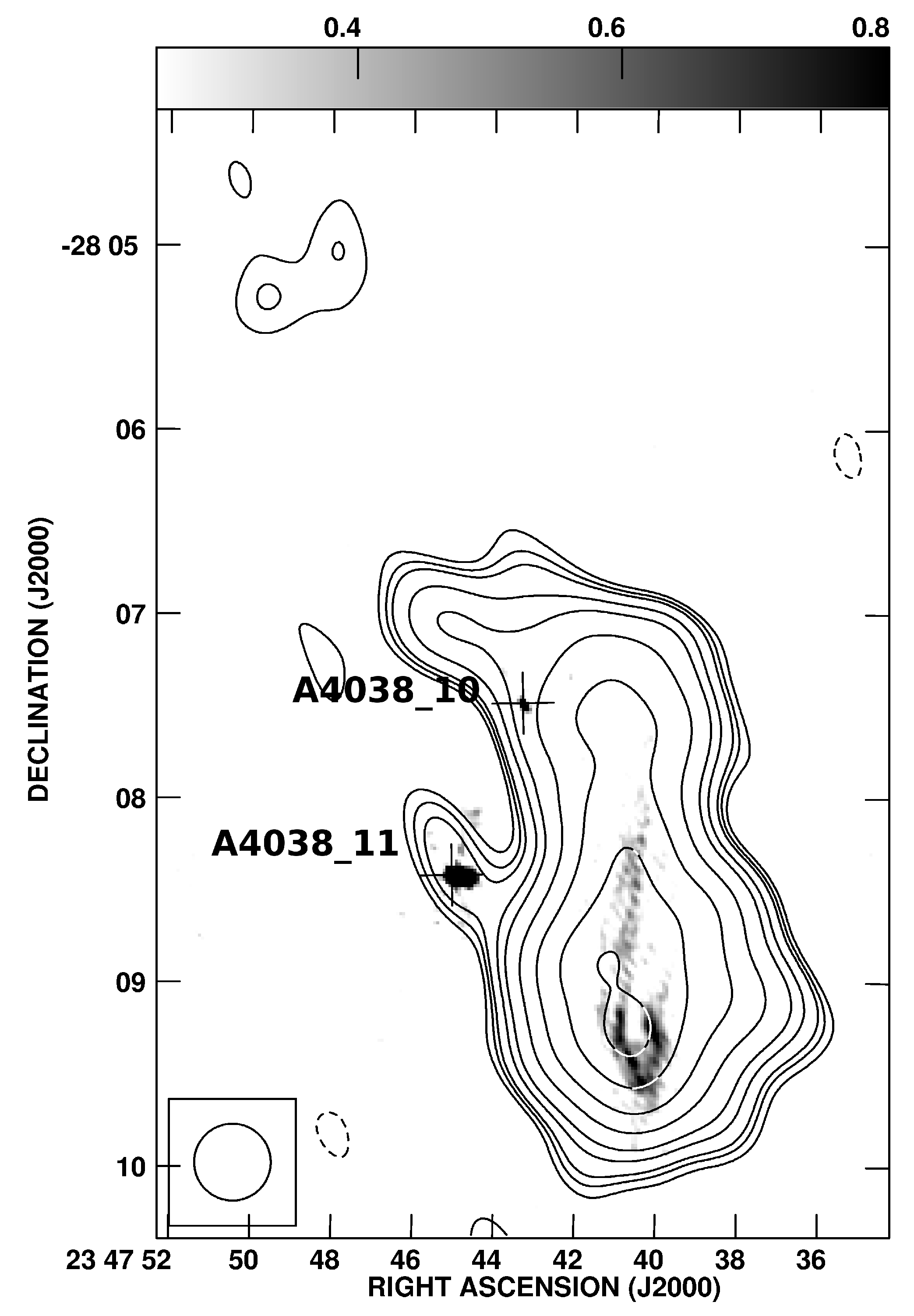}
 \caption[A4038 GMRT 150 and 1288 MHz images]{A4038: GMRT 150 MHz ($25''\times25''$) contours overlaid on GMRT 1288 MHz ($3.8''\times2.2''$, Position angle$=36^\circ$) image shown in grey- scale. The contours are at -15, 15, 20, 25, 40, 60, 120, 240, 480, 770 mJy beam$^{-1}$. Grey-scale extends from 0.25 to 0.8 \mjyb. The plus signs mark the positions of discrete radio sources A4038$\_$10 and A4038$\_$11. \label{fig:Fig. 1}}
\end{figure}

\begin{figure}
\includegraphics[width=7 cm]{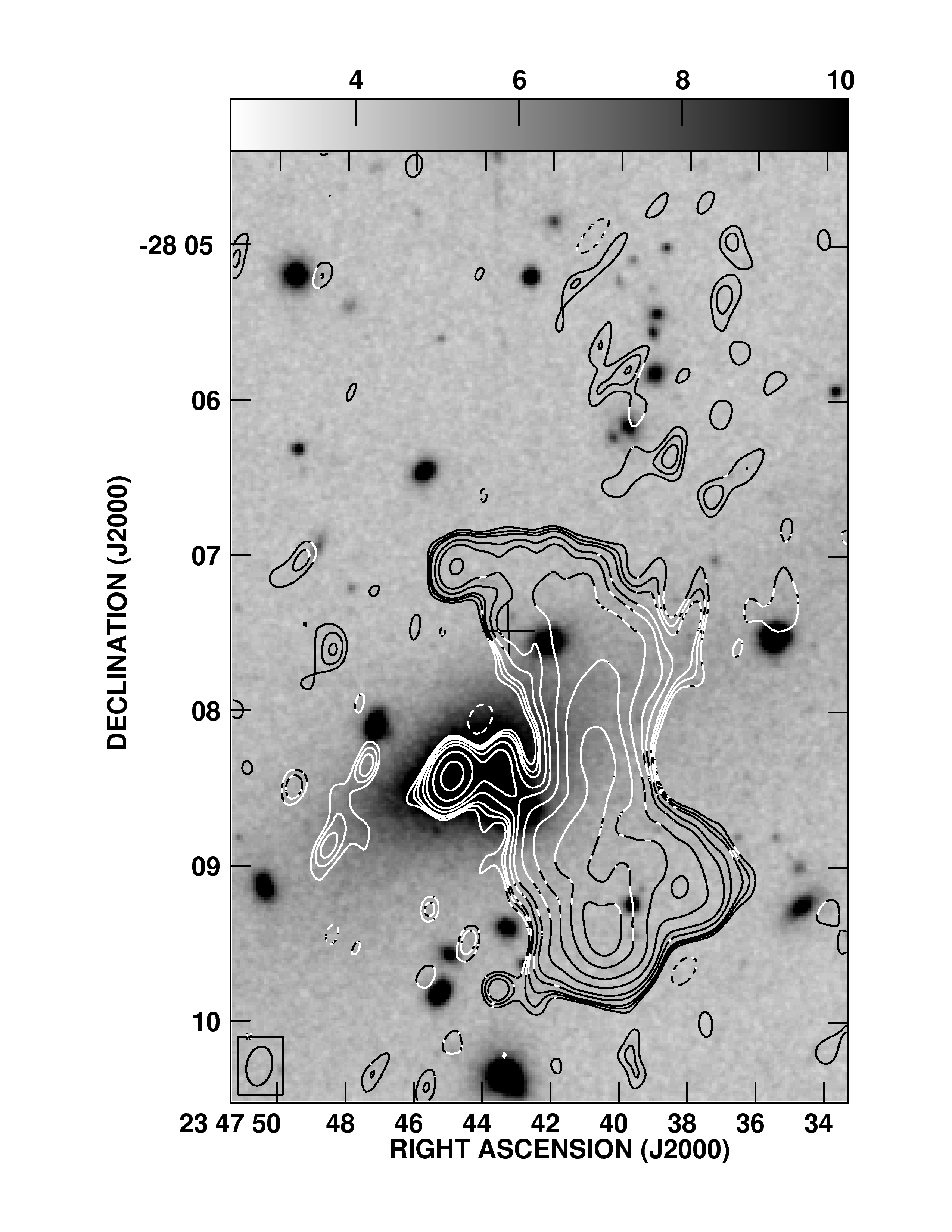}
\includegraphics[width=7 cm]{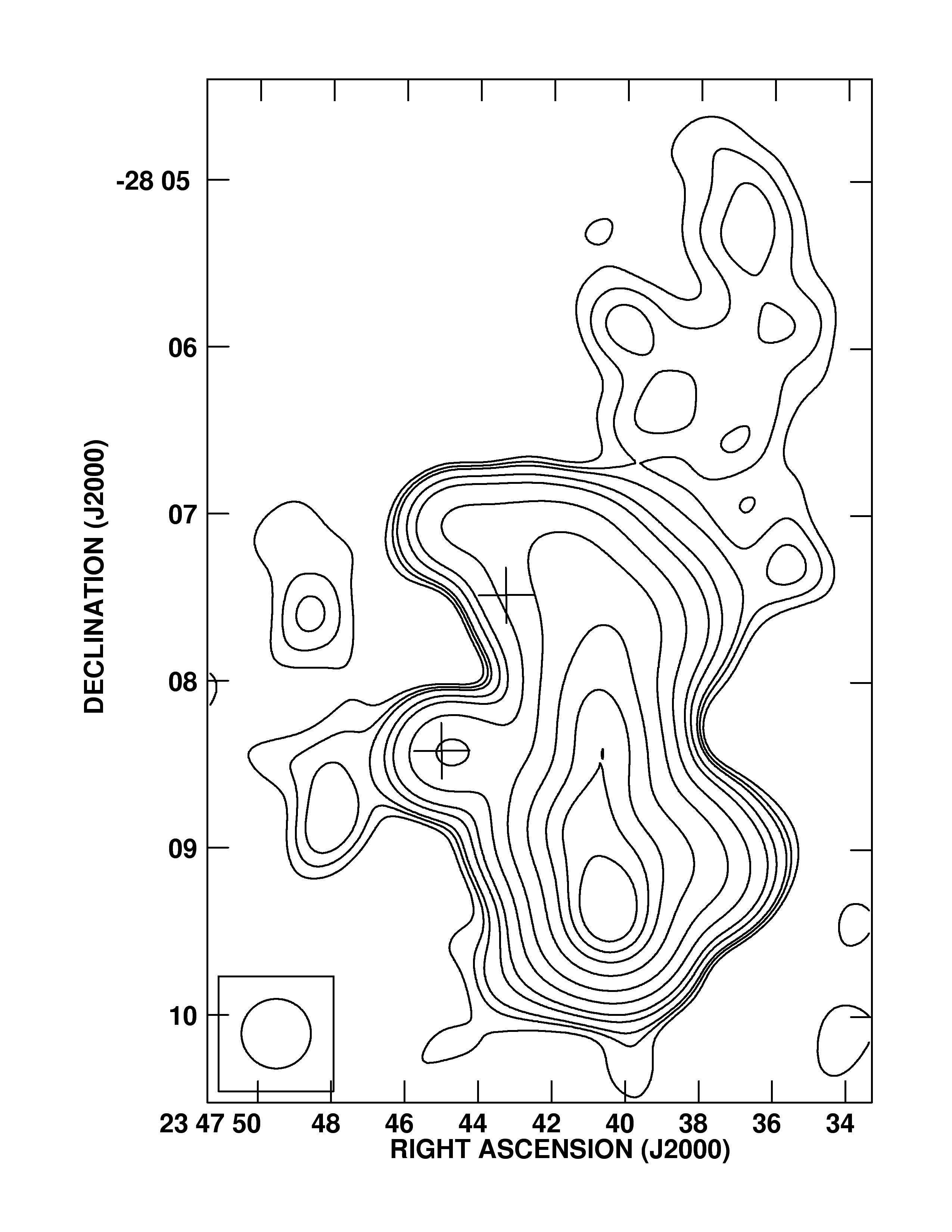}
 \caption{A4038: {\it Left:} GMRT 240 MHz ($15''\times10''$, Position angle$= -11^\circ$) contours overlaid 
on optical R band image (DSS) shown in grey- scale. The contours are at -2.7, 2.7, 3.6, 4.5, 7.2, 
10.8, 21.6, 43.2, 86.4, 138.6 mJy beam$^{-1}$. The grey-scale extends from 2500 to 10000 counts. {\it Right:} GMRT 240 MHz image ($25''\times25''$). The contours are at -4.5, 4.5, 6.0, 7.5, 12.0, 18.0, 36.0,72.0, 144.0, 231.0 mJy beam$^{-1}$.\label{fig:Fig. 2}}
\end{figure}

\begin{figure}
 \epsscale{0.5}
  \plotone{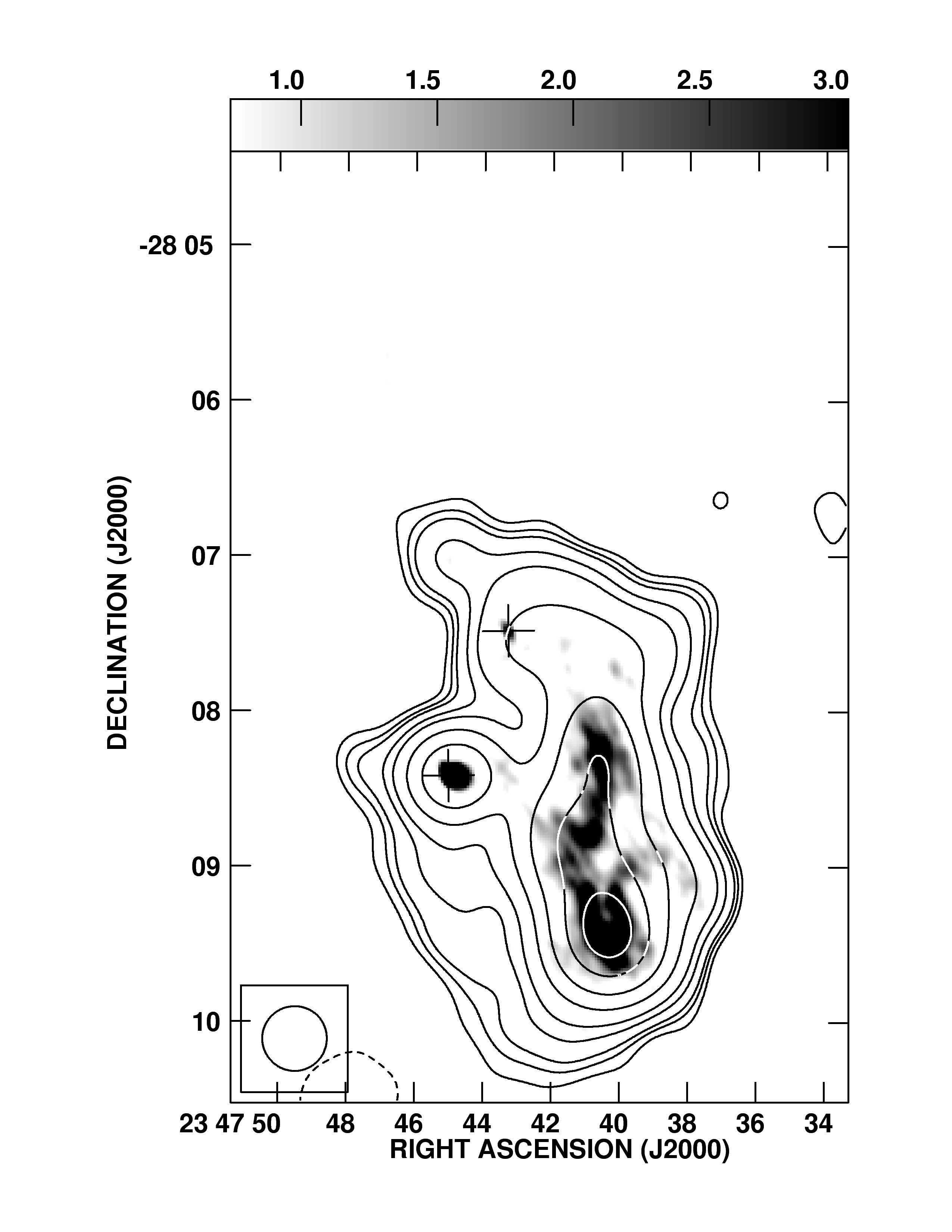}
 \caption[]{A4038: GMRT 606 MHz ($25''\times25''$) low resolution image shown in contours overlaid on GMRT 
606 MHz ($6.6''\times3.5''$, Position angle$=  28^\circ$) high resolution image in grey-scale. The contours are at -1.2, 1.2, 1.6, 2.0, 3.2, 4.8, 9.6, 19.2, 38.4, 61.6 mJy beam$^{-1}$. The grey-scale extends over the range of 0.75 to 3 \mjyb. \label{fig:Fig. 3}}
\end{figure}

\begin{figure}
	\centering	
		\includegraphics[width=7.1 cm]{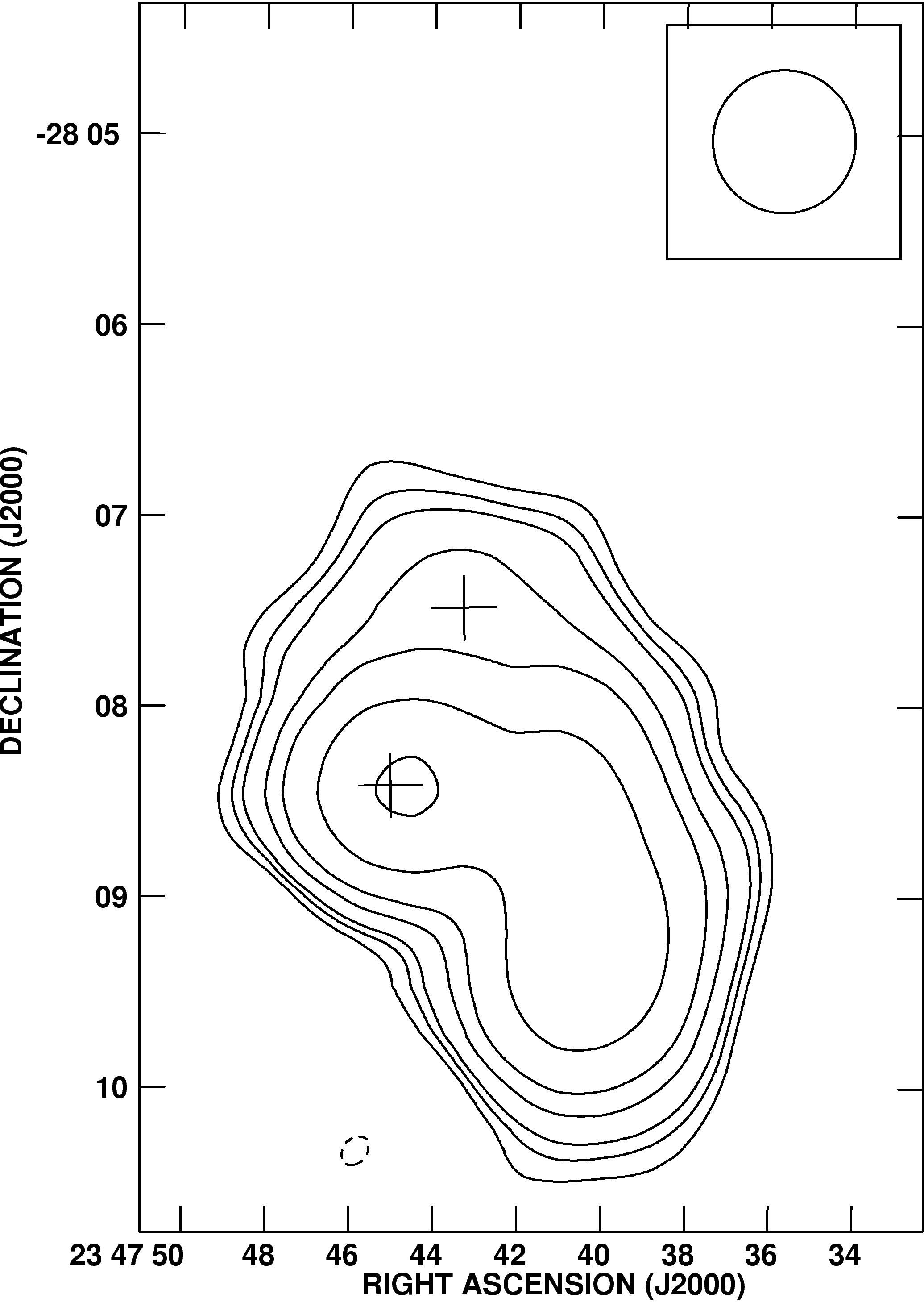}
		\includegraphics[width=7.1 cm]{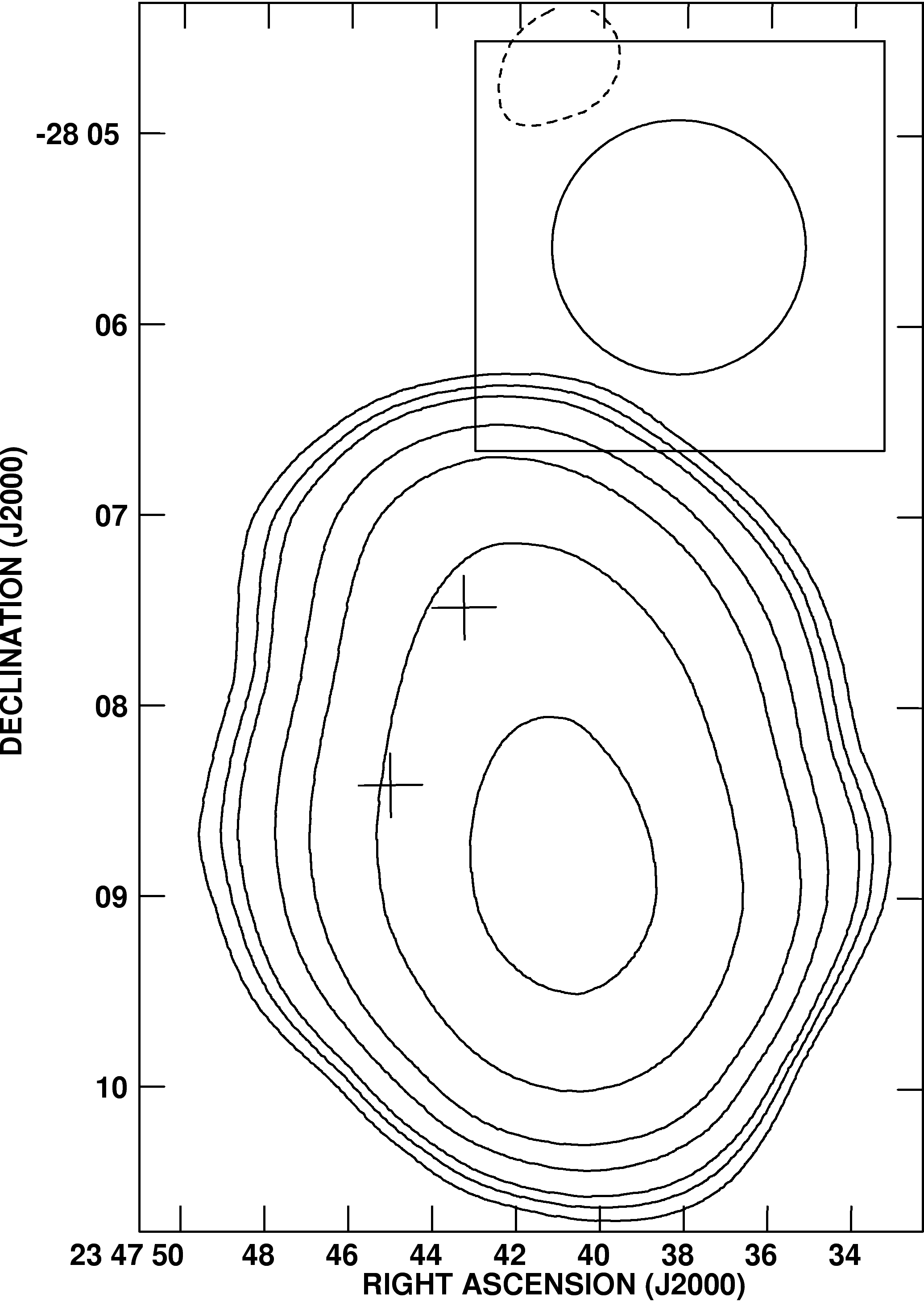}
\caption[Images of A4038 at 1400 and 74 MHz]{A4038: {\it Left:} Image of A4038 from the NVSS 
(1400 MHz). The synthesized beam is $45''\times45''$. {\it Right:} 
VLSS (74 MHz) image of A4038. The synthesized beam is	$80''\times80''$. 
Contours are at (left) -1.35, 1.35, 1.80, 2.25, 3.6, 5.4, 10.8, 
21.6 mJy beam$^{-1}$ and at (right) -300, 300, 400, 500, 800, 1200, 2400, 4800
 mJy beam$^{-1}$.}\label{fig:Fig. 4}
\end{figure}

\begin{figure}
	\centering	
		\includegraphics[width=8.0cm]{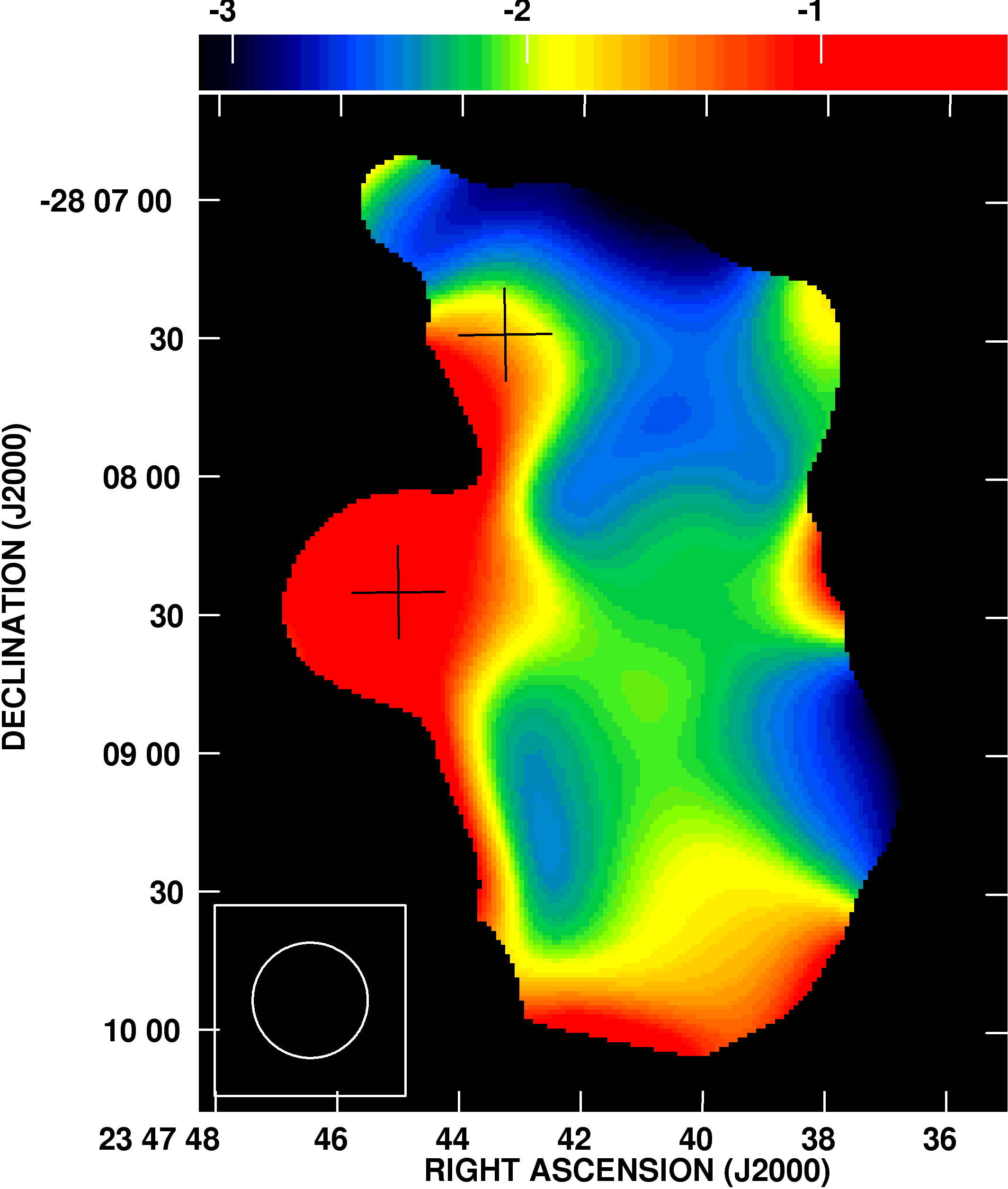}
		\includegraphics[width=8.0cm]{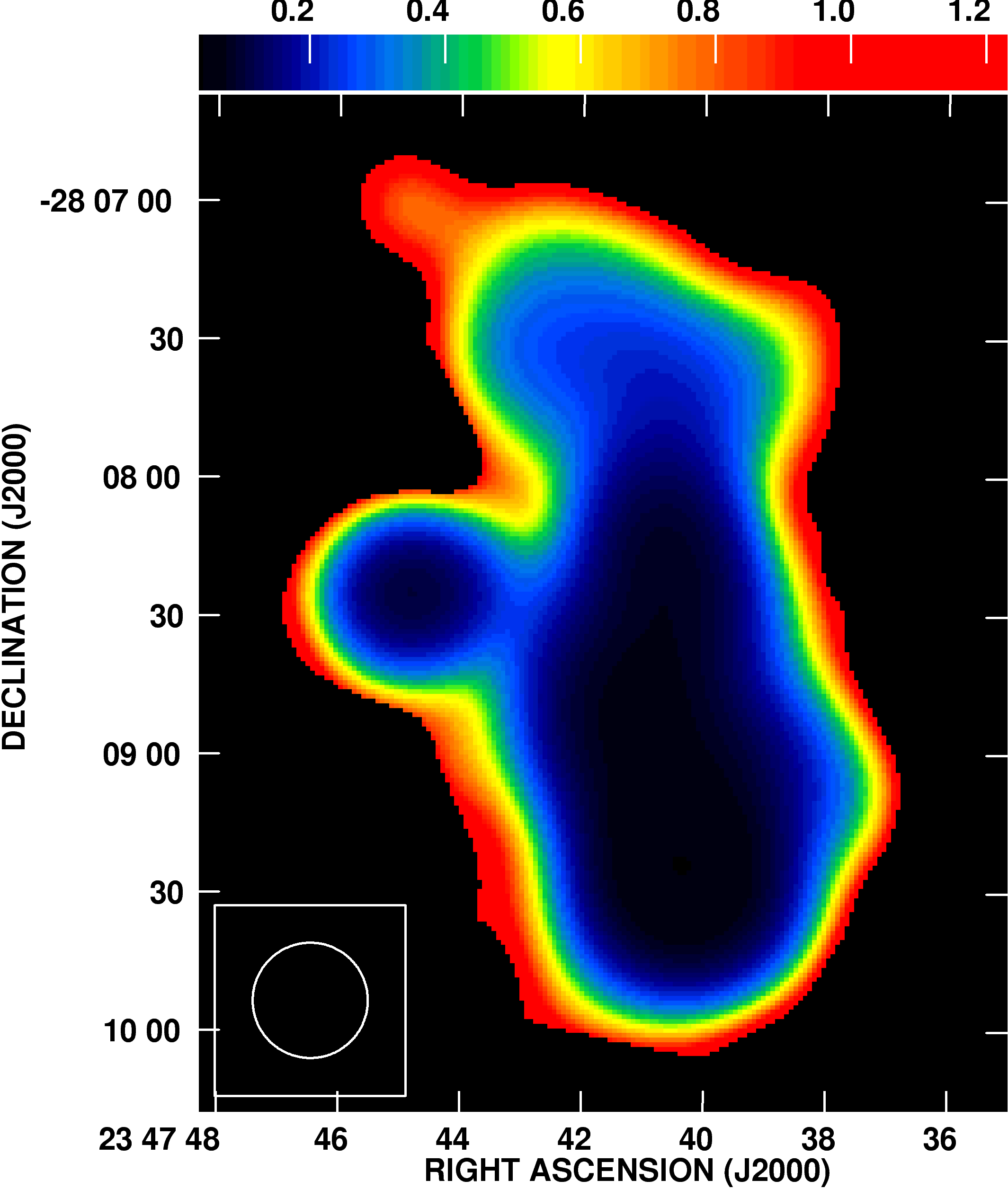}	
\caption[Spectral index images of A4038 between 606 and 240 MHz]{A4038: {\it Left:} Spectral index 
image of A4038 between 606 and 240 MHz. The synthesized beam is $25''\times25''$. {\it Right:} Spectral index uncertainty image.} \label{fig:Fig. 5}
\end{figure}

\begin{figure}
	\centering
		\includegraphics[width=5.1cm,angle=-90]{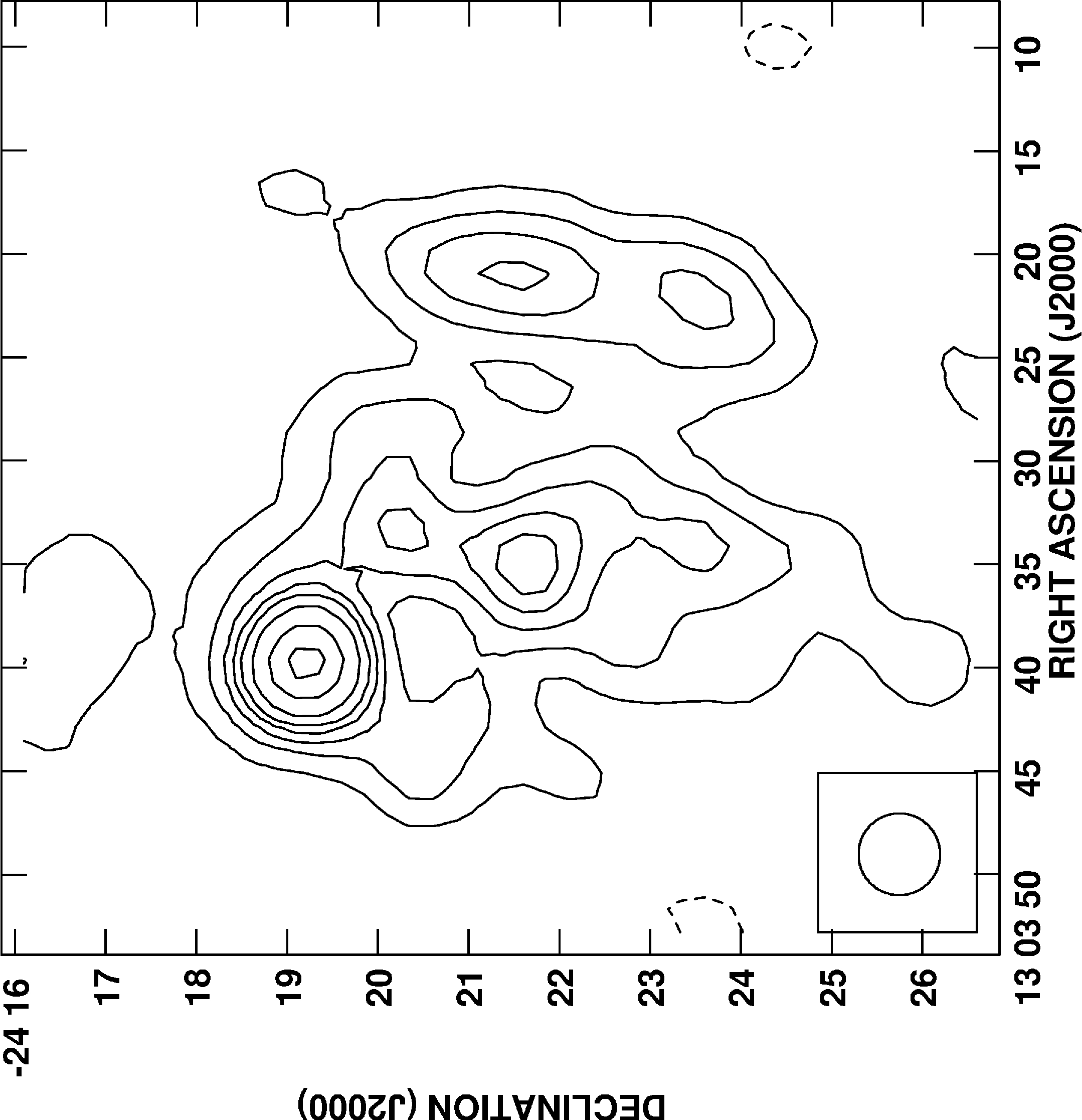}
		\includegraphics[width=5.1cm,angle=-90]{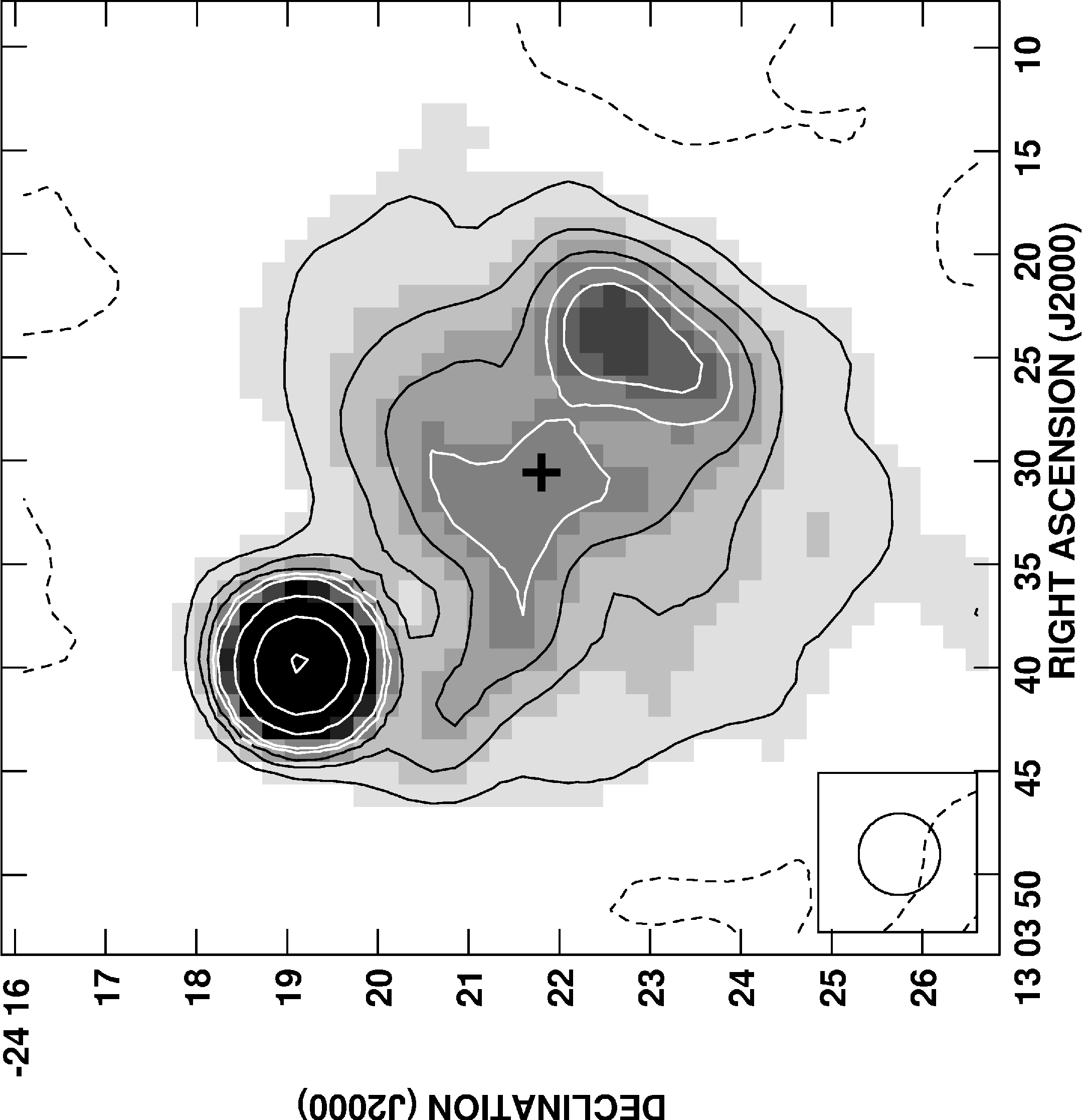}
		\includegraphics[width=5.1cm,angle=-90]{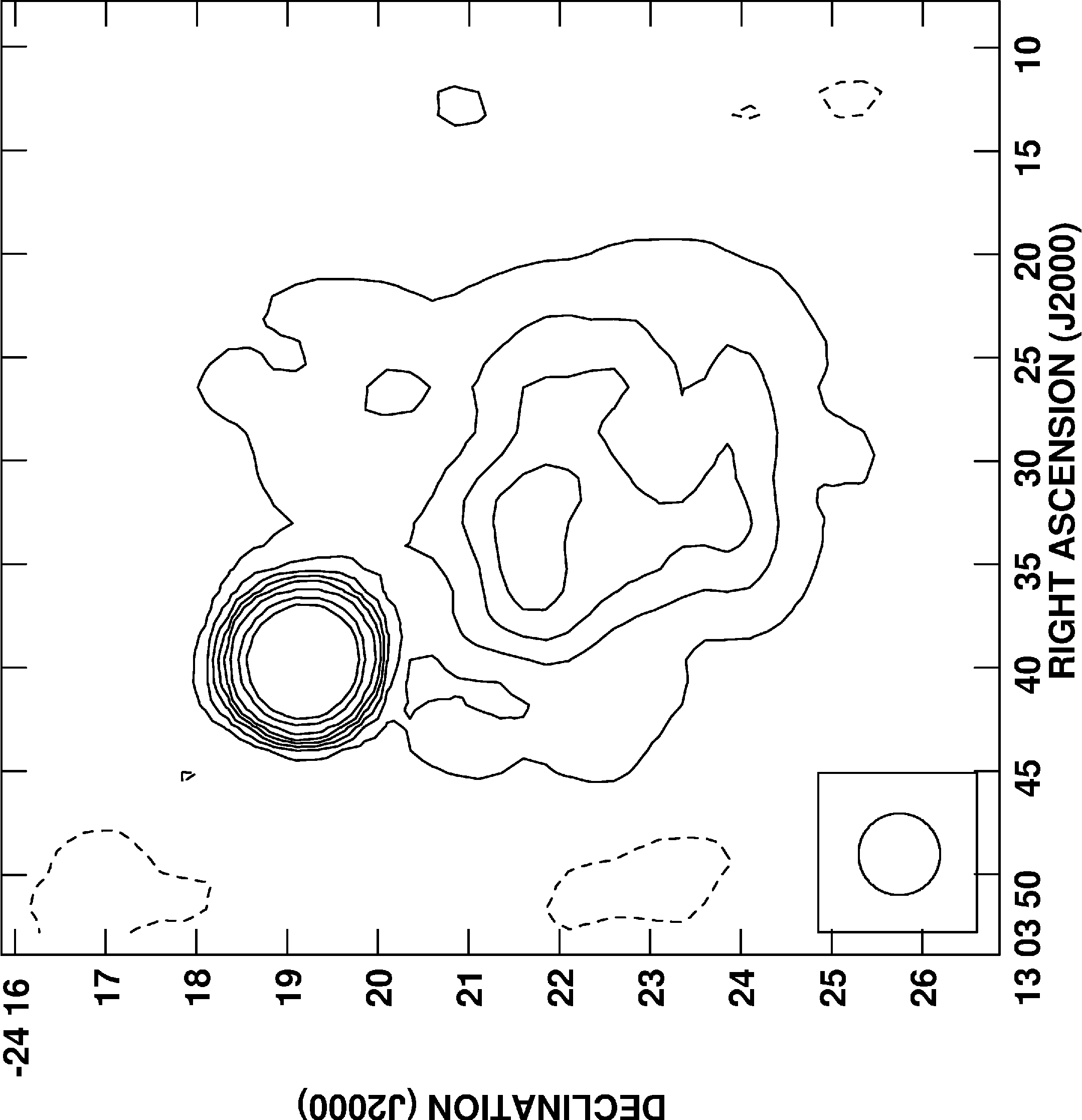}	
\caption[Images of A1664 at 150, 325 and 1400 MHz]{A1664: {\it Left:} GMRT 150 MHz 
contours. Contour levels are at -21, 21, 42, 63, 84, 105, 210, 420, 840 \mjyb (synthesized beam $= 54''\times54''$). {\it Center:} GMRT 325 MHz contour levels at -4.5, 4.5, 9.0, 13.5, 18.0, 22.5, 45.0, 90.0, 180.0 \mjyb (synthesized beam $=54''\times54''$). Grey-scale extends from 3 to 40 \mjyb. The location of the unresolved source detected in the high resolution image at 325 MHz is marked ($'+'$). {\it Right:} NVSS 1400 MHz contour levels at -1.05, 1.05, 2.10, 3.15, 4.20, 5.25, 10.50, 21.00, 42.00 \mjyb (synthesized beam $= 45''\times45''$). The source A$1664\_3$ is seen at the north of the relic.}
\label{fig:Fig. 6}
\end{figure}

\begin{figure}
\includegraphics[height=8.0cm,angle=-90]{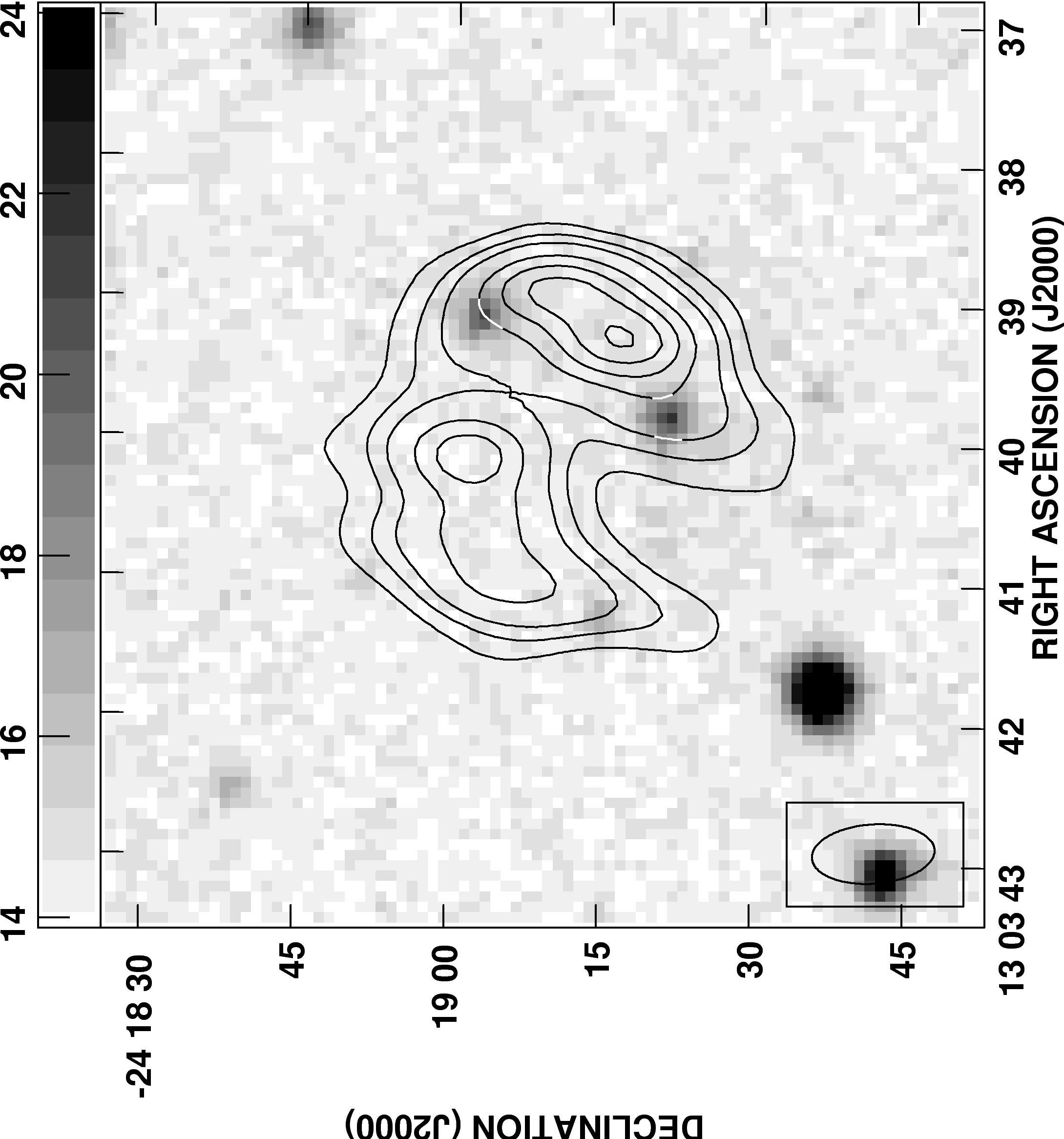}
\caption[A1664 compact source resolved]{A1664: The source A1664$\_3$ as seen at 325 MHz with high resolution (synthesized beam$=12''\times6''$, Position angle$= 5^\circ$) . 
The 325 MHz contours (-3, 3, 6, 9, 15, 20, 25, 30, 30, 35, 40 \mjyb) are overlaid on the DSS R band image of the same field shown in grey-scale.
 \label{fig:Fig. 7}}
\end{figure}

\begin{figure}
	\centering	
		\includegraphics[width=8.0cm]{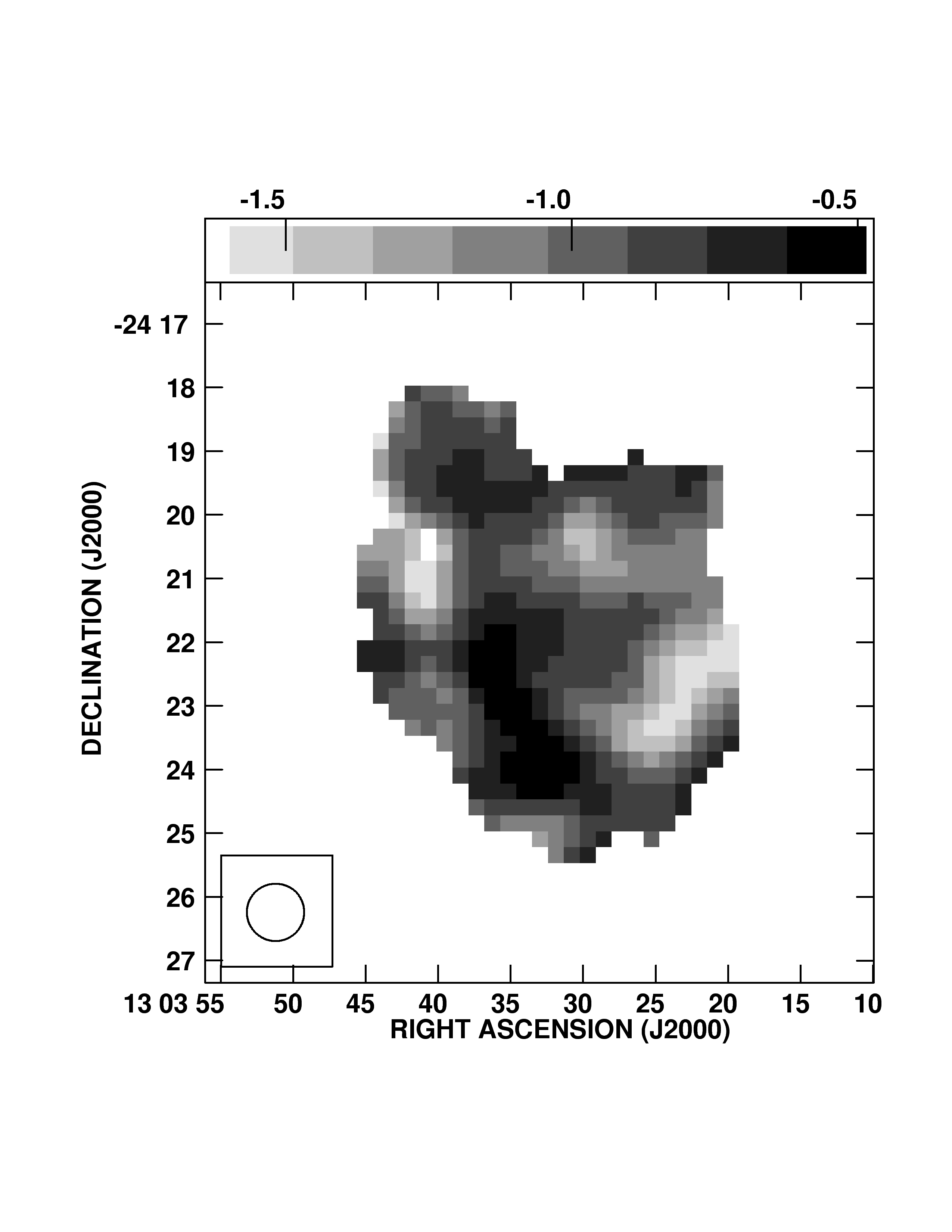}
		\includegraphics[width=8.0cm]{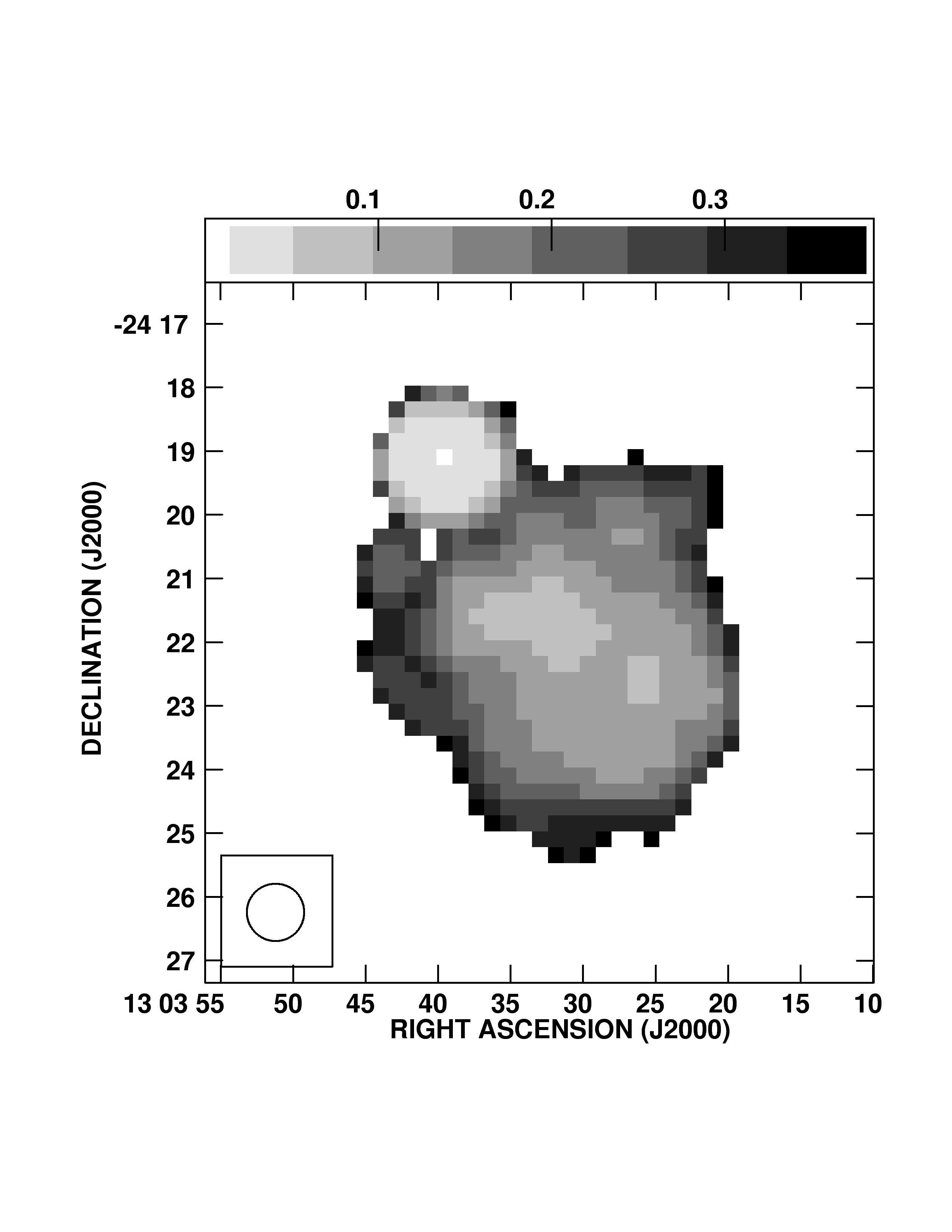}
\caption[Spectral index images of A1664 between 1400 and 325 MHz]{A1664: {\it Left:} Spectral index image of A1664 between 1400 and 325 MHz. The synthesized beam is $25''\times25''$. {\it Right:} Spectral index uncertainty image.\label{fig:Fig. 8}}
\end{figure}

\begin{figure}
	\centering	
\includegraphics[width=6.0 cm,angle=-90]{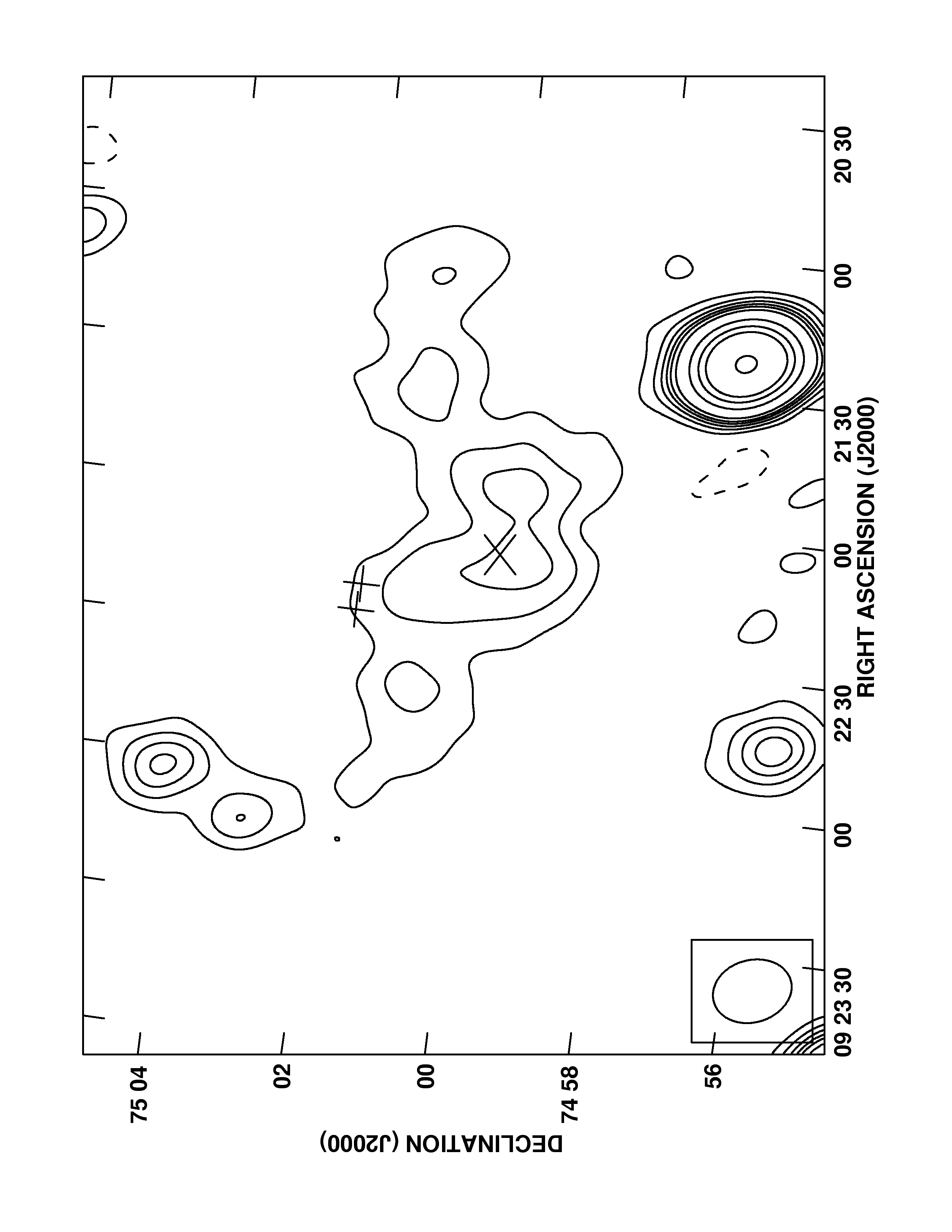}
\includegraphics[width=6.0 cm,angle=-90]{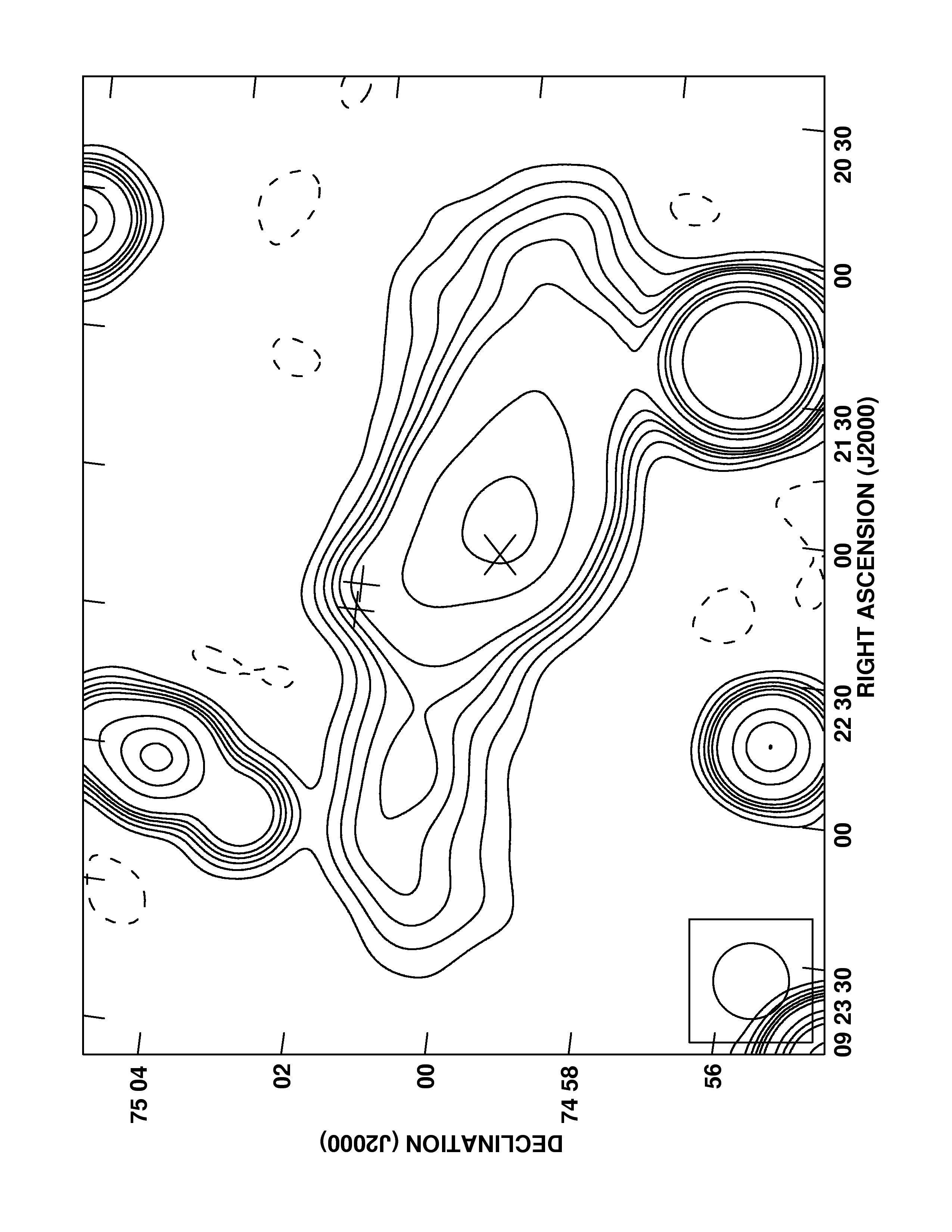}
\includegraphics[width=6.0 cm,angle=-90]{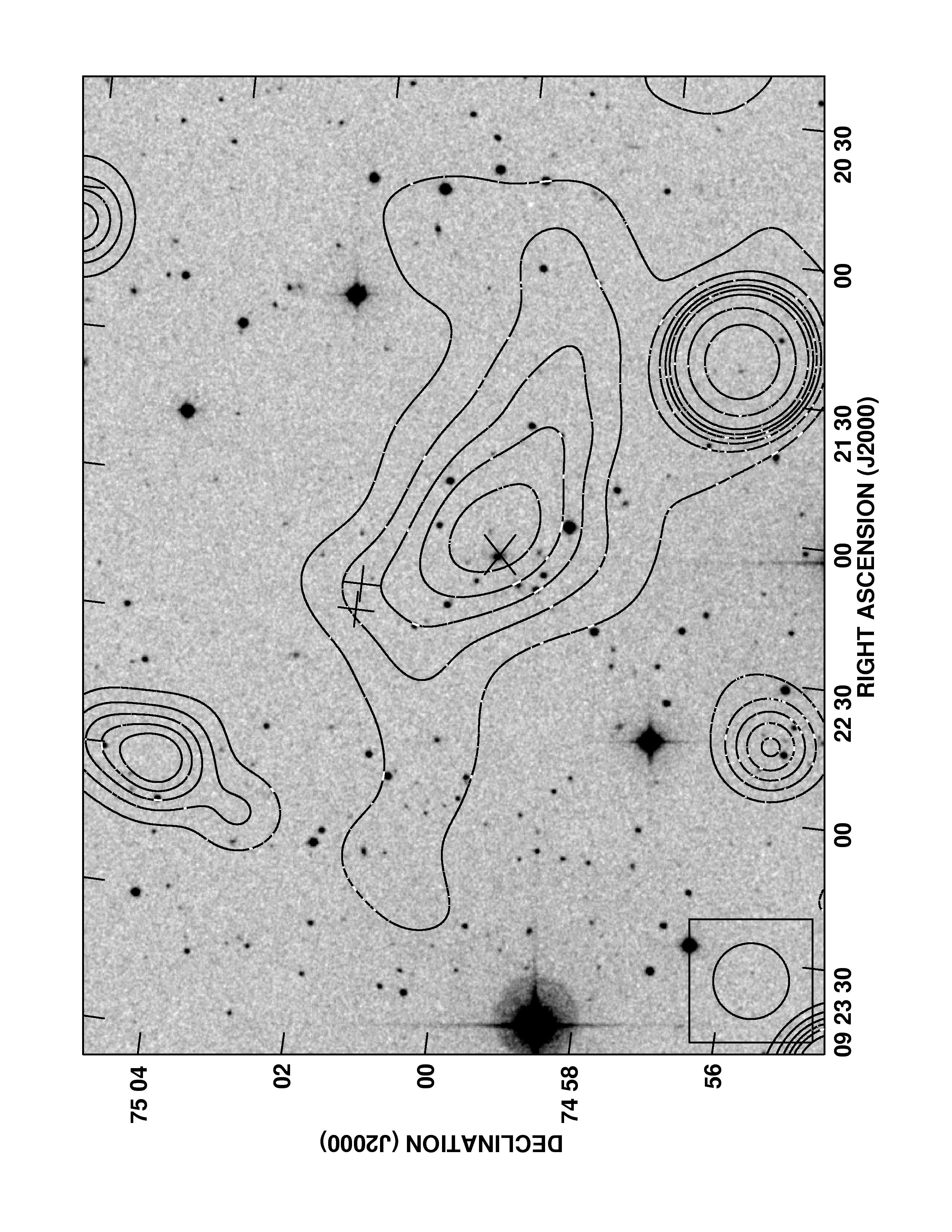}
\includegraphics[width=6.0 cm,angle=-90]{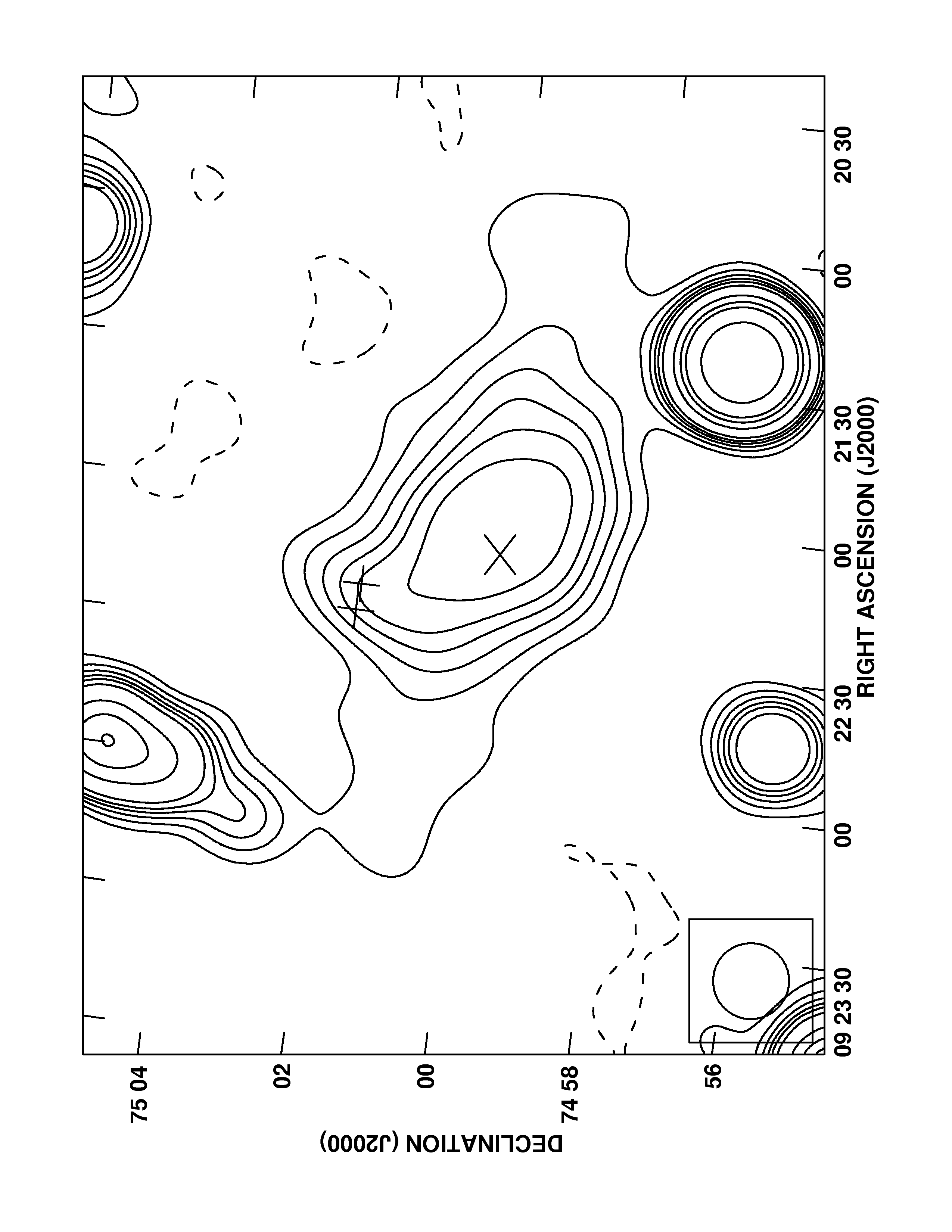}
\caption[Images of the relic near A786 at 150, 345, 606 and 1400 MHz]
{A786: {\it Top Left:} GMRT 150 MHz. Contour levels are at -30, 30, 60, 90, 120, 150, 200, 400, 600, 800, 1600 \mjyb (synthesized beam$=67''\times52''$, P.A.$=22^\circ$). {\it Top Right:} WSRT 345 MHz. 
Contour levels are at -2.4, 2.4, 4.8, 7.2, 9.6, 12.0, 16.0, 32.0, 48.0, 64.0, 128.0 \mjyb (synthesized beam$=64''\times64''$).
{\it Bottom Left:} GMRT 606 MHz contours overlaid on DSS R band image in grey-scale. Contour levels are at -6.9, 6.9, 20.7, 27.6, 34.5, 46.0, 92.0, 184.0, 368.0 \mjyb (synthesized beam$=64''\times64''$). {\it Bottom Right:} VLA 1400 MHz. Contour levels are at -0.3, 0.3, 0.6, 0.9, 1.2, 1.8, 2.4, 3.0, 4.0, 5.0, 7.0, 10.0, 12.0, 20.0, 50.0 \mjyb (synthesized beam $=64''\times64''$).
\label{fig:Fig. 9}}
\end{figure}

\begin{figure}
	\centering
	\includegraphics[width=8 cm, angle =0]{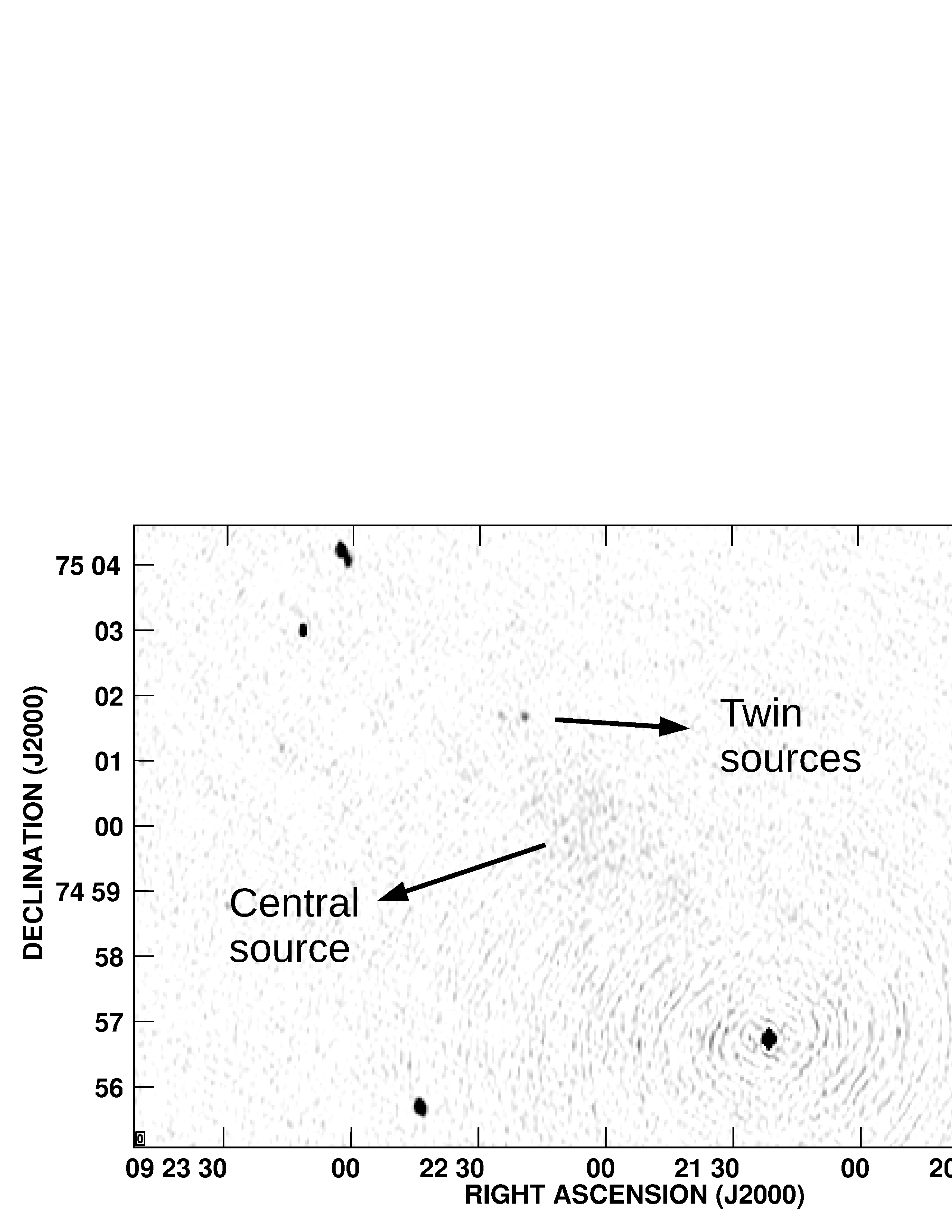}
	\includegraphics[width=8 cm, angle =0]{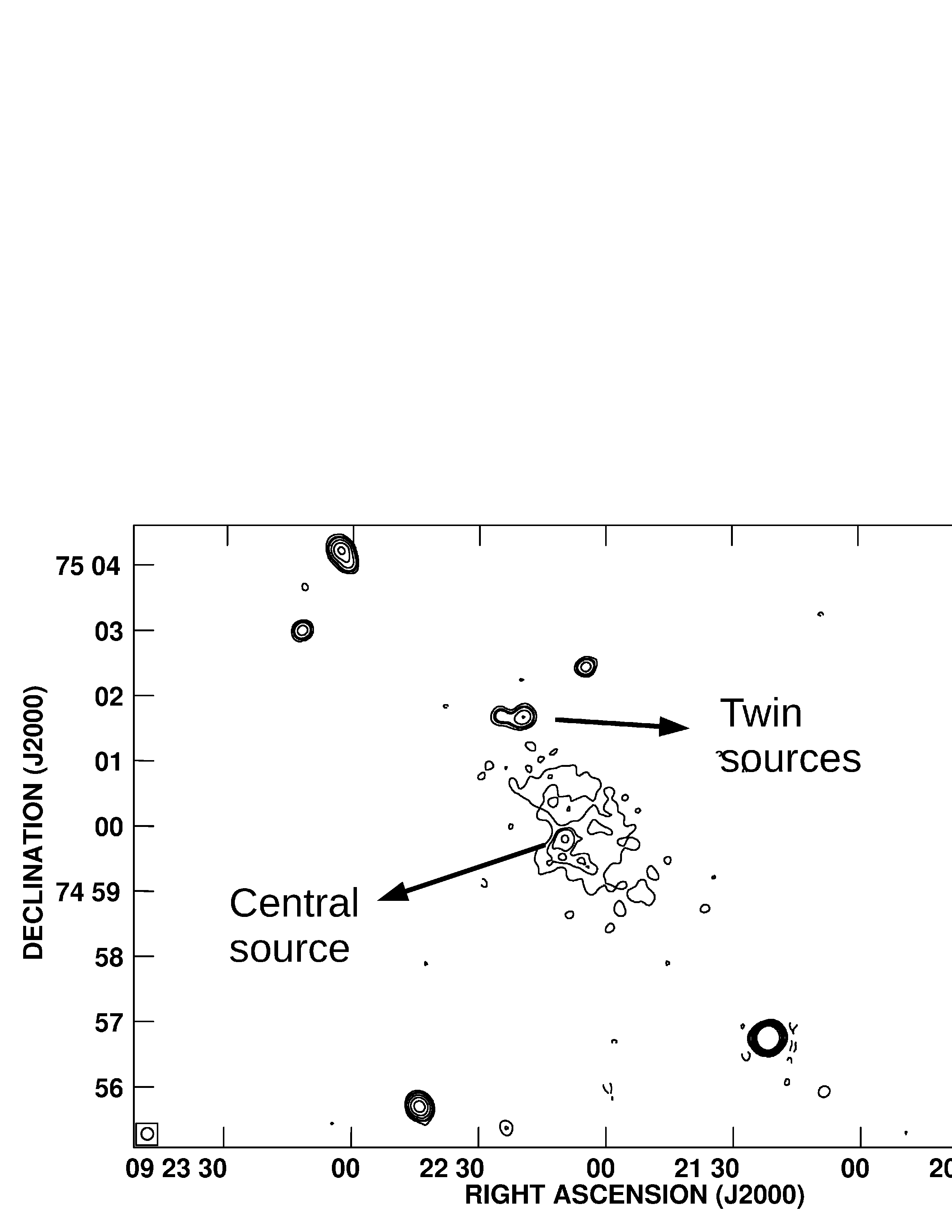}
\caption[Image of the relic near A786 at 4.7 GHz]{A786: Images of the region of the relic in A786 at 606 MHz (left) and 4.7 GHz (right). A central radio source with extended emission surrounding it is detected at 4.7 GHz. The twin sources are detected at 606 MHz but the central source is not. The synthesized beams are  $7''\times4''$ (P. A.$=0.16^\circ$) and $11''\times11''$ (P. A.$=-64^\circ$) in the left and in the right panels respectively.\label{fig:Fig. 10}}
\end{figure}

\begin{figure}
	\centering	
	\includegraphics[width=6.0cm,angle=-90]{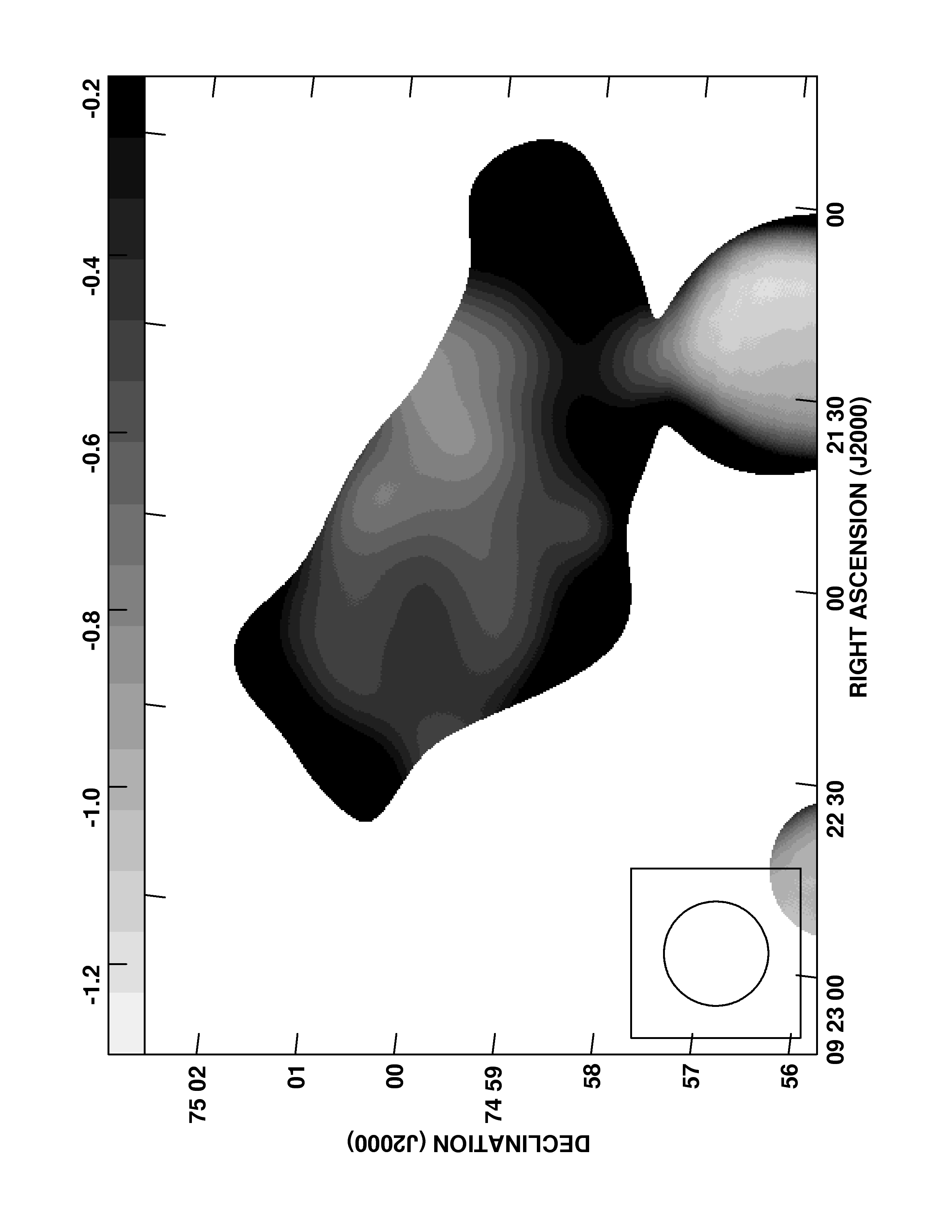}
	\includegraphics[width=6.0cm,angle=-90]{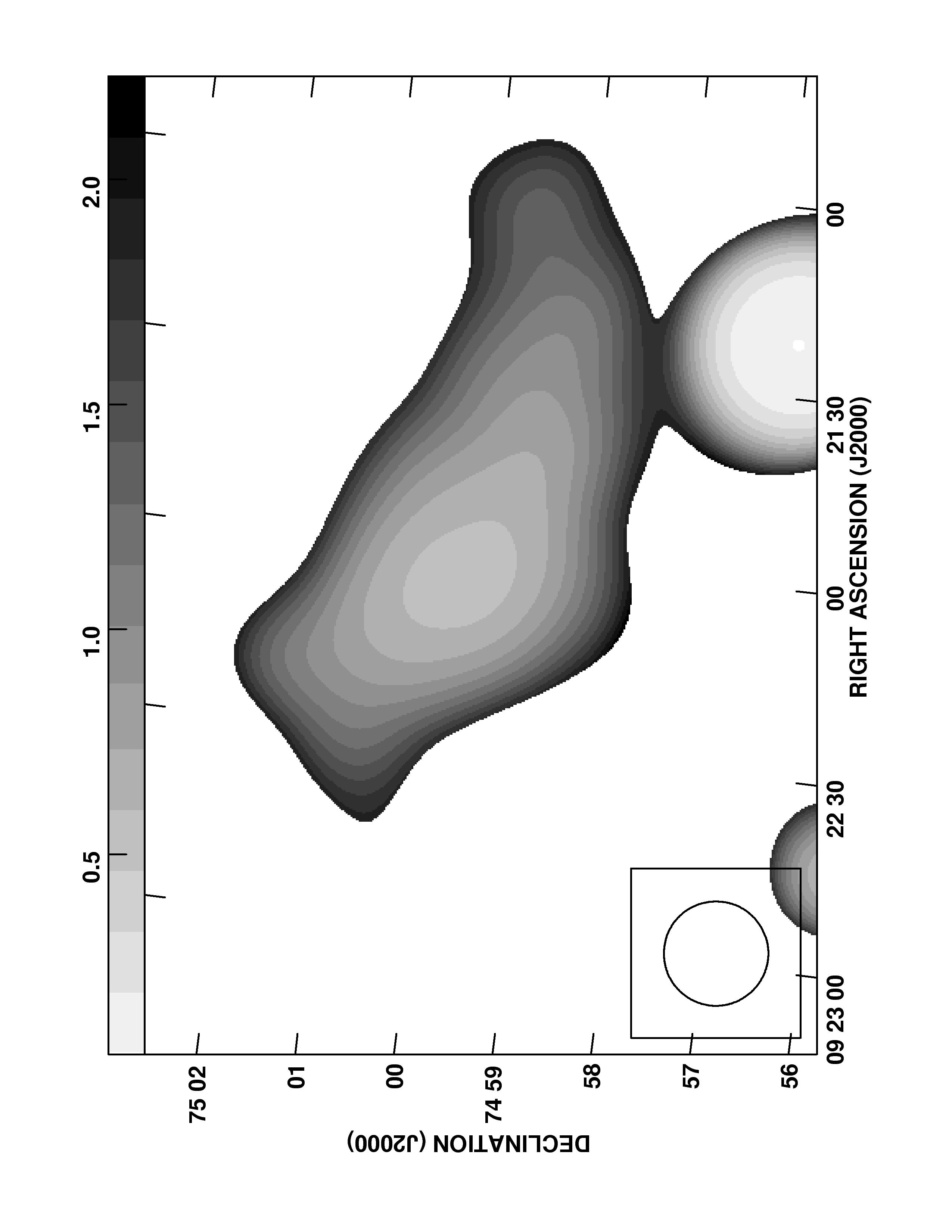}
\caption[Spectral index images of A786 relic between 606 and 345 MHz]{A786: {\it Left:} Spectral 
index image of A786 between 606 and 345 MHz. The synthesized beam is $64''\times64''$. {\it Right:} Spectral index uncertainty image.}
\label{fig:Fig. 11}
\end{figure}
\begin{figure}
	\centering	
		\includegraphics[width=6.0cm,angle=-90]{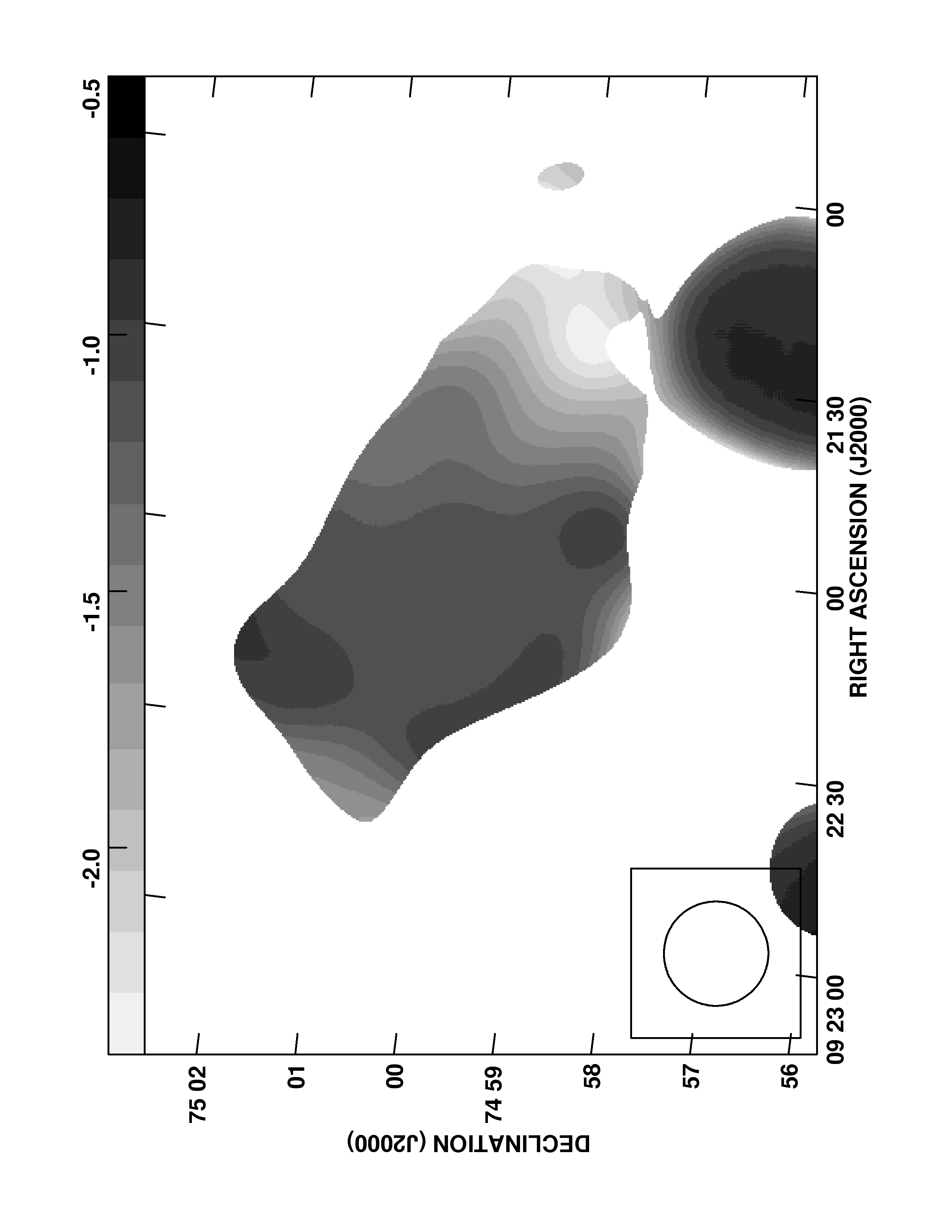}
		\includegraphics[width=6.0cm,angle=-90]{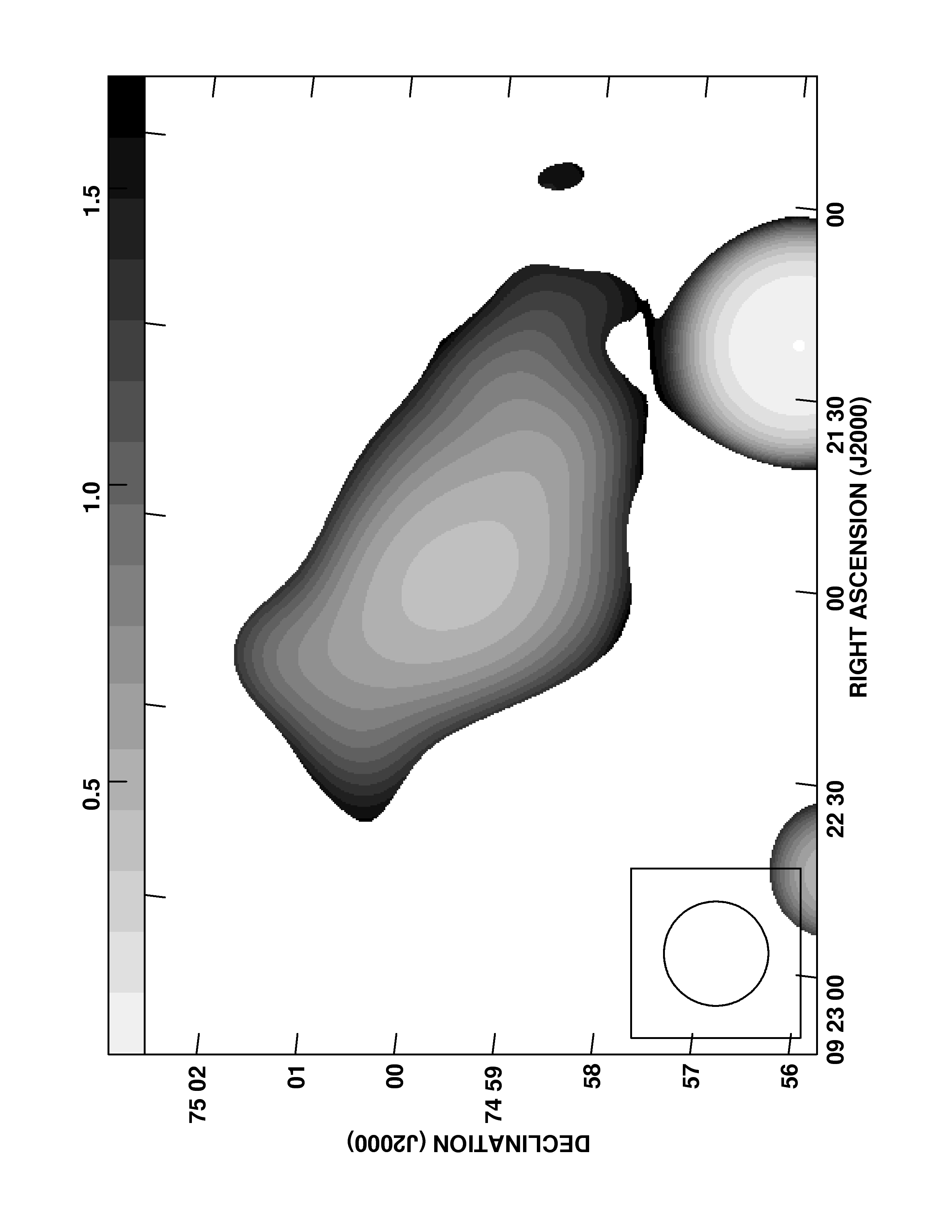}
\caption[Spectral index images of A786 relic between 606 and 1400 MHz]{A786: {\it Left:} Spectral 
index image of A786 between 606 and 1400 MHz. The synthesized beam is $64''\times64''$. {\it Right:} Spectral index uncertainty image.}
\label{fig:Fig. 12}
\end{figure}

\begin{figure}
	\includegraphics[width=7.2 cm]{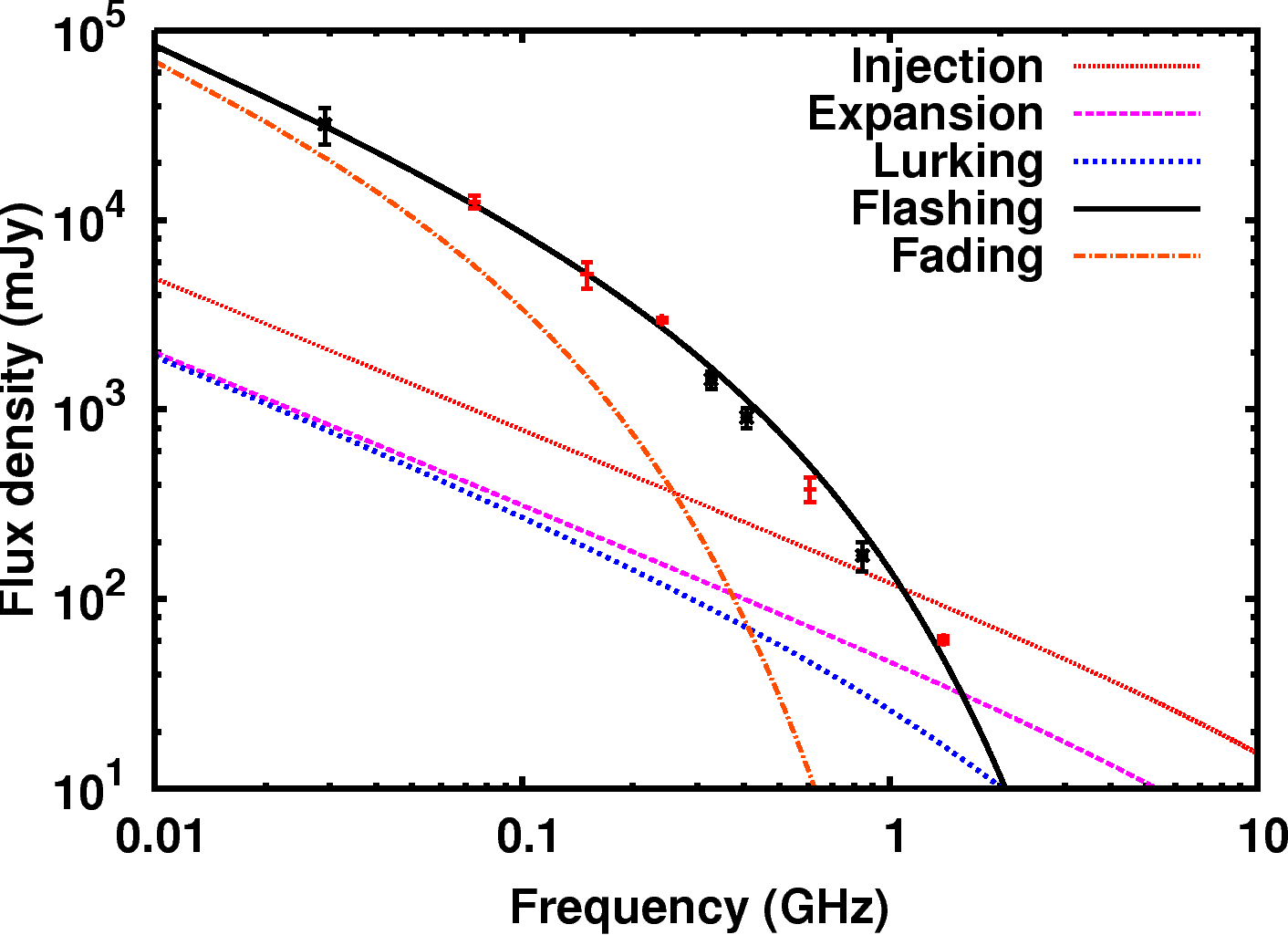}
	\includegraphics[width=7.2 cm]{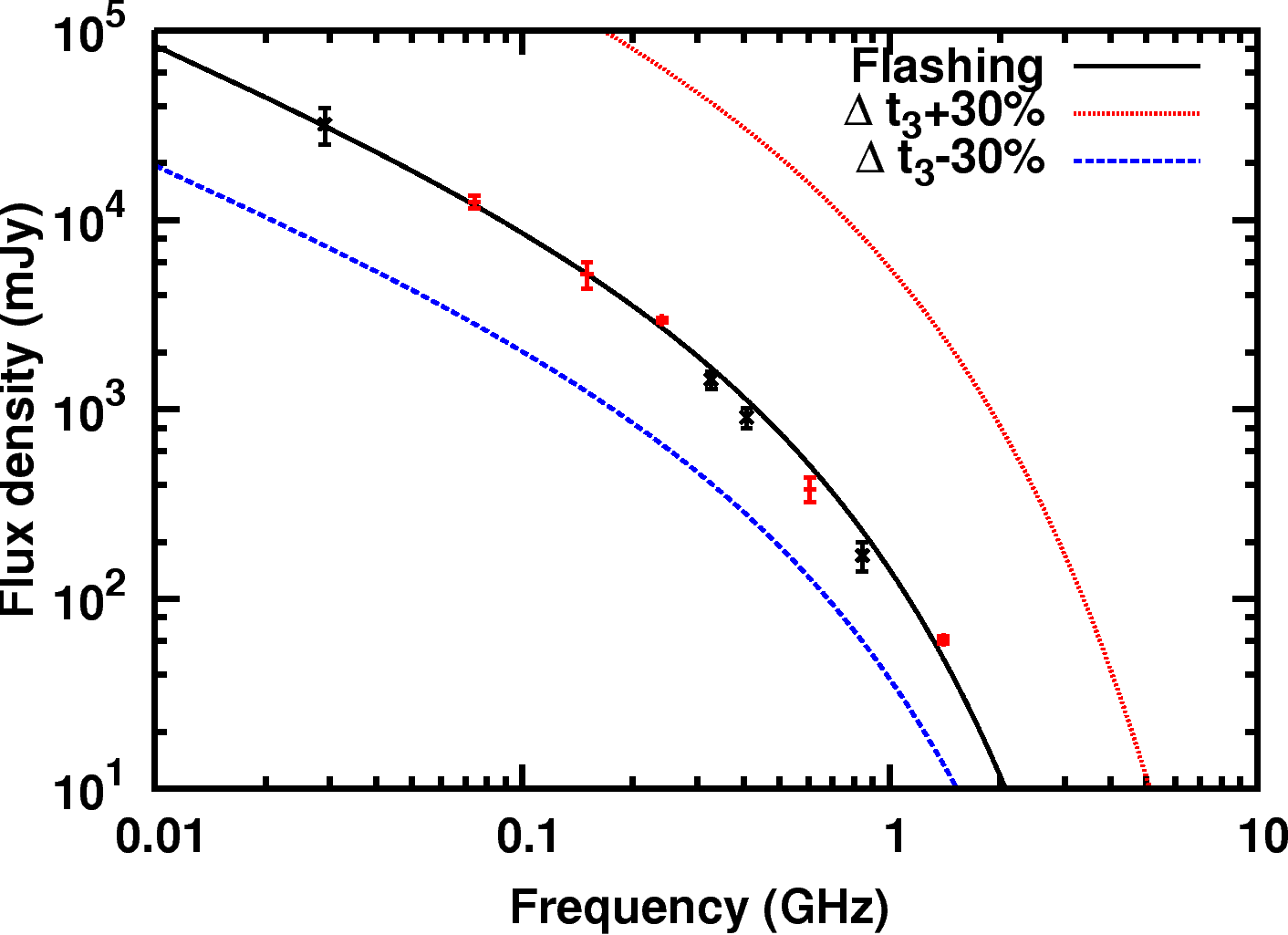}
\caption[Model fit in `flashing' phase for A4038]{A4038: {\it Left:} The best fit model 
spectrum in `flashing' phase along with spectra in other phases. {\it Right:} The 
observed integrated spectrum of the relic with the best fit and the model spectra with 
$\Delta t_{3}\pm30\%$. The red points mark values estimated using our GMRT images (150, 240 and 606 MHz) and the images from the VLSS (74 MHz) and the NVSS (1400 MHz). Black points mark values 
from Slee et al (2001). See the electronic edition of the Journal for a color version 
of this figure.\label{fig:Fig. 13}}
\end{figure}

\begin{figure}
	\includegraphics[width=7.2 cm]{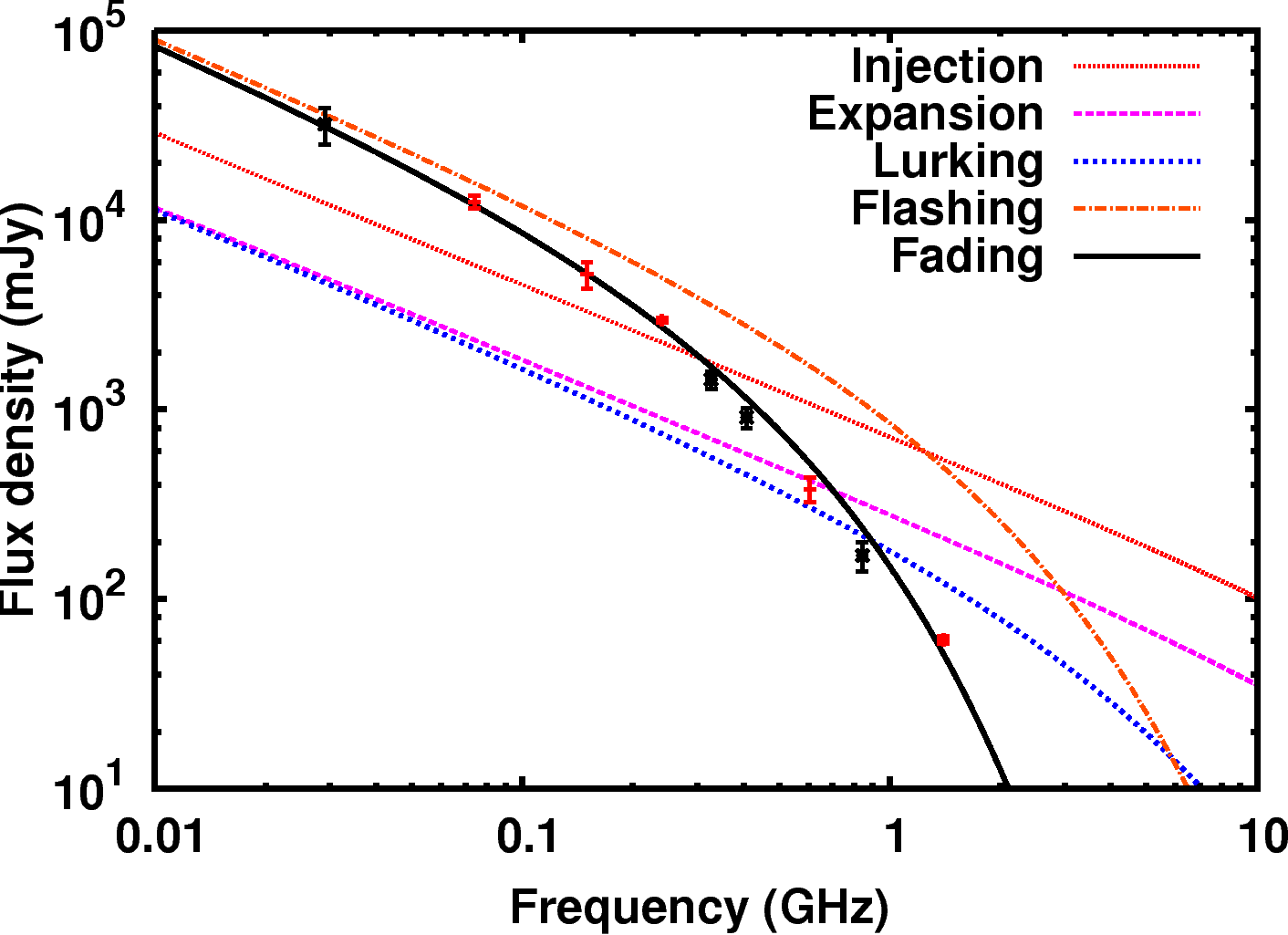}
	\includegraphics[width=7.2 cm]{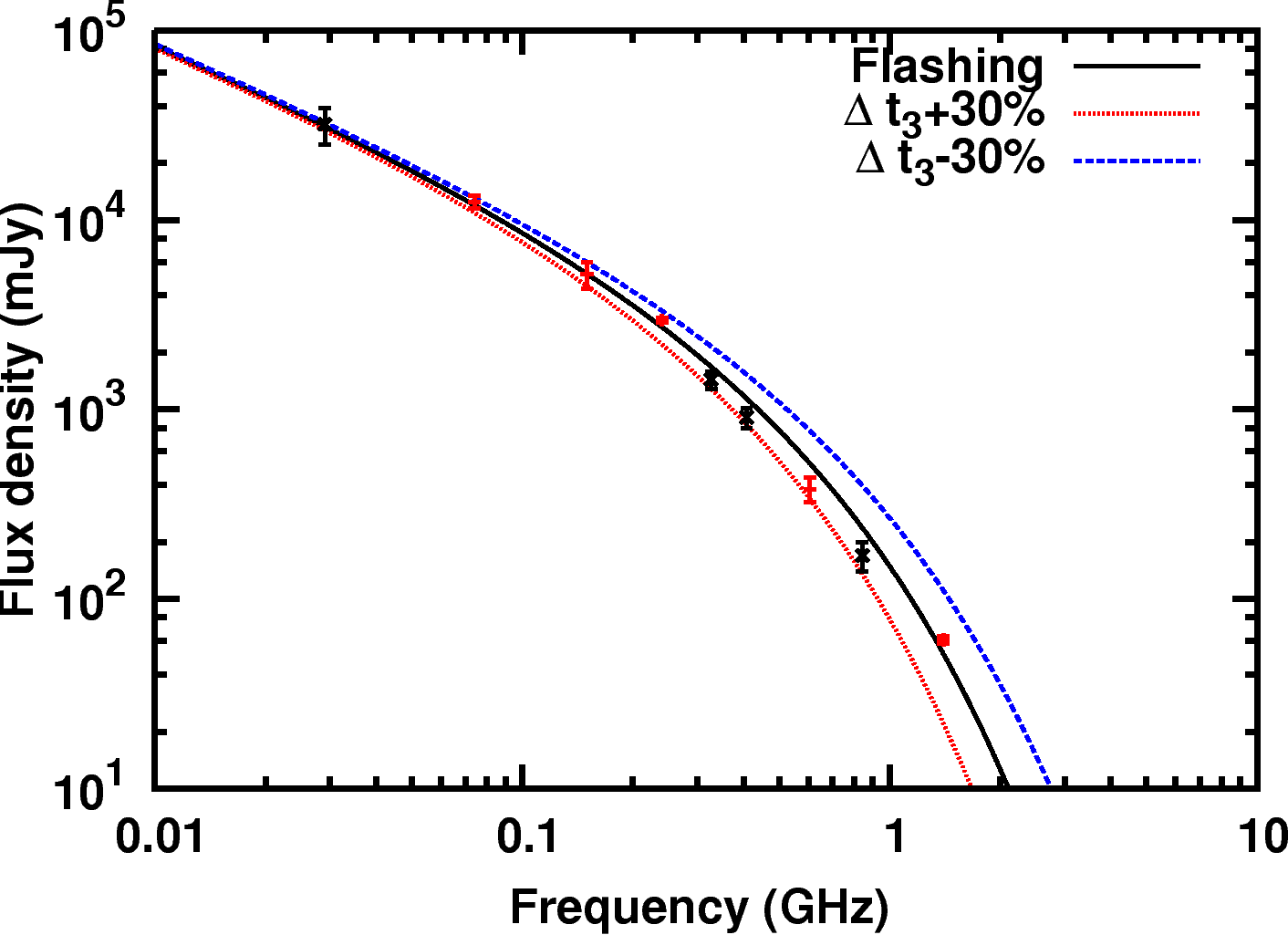}
\caption[Model fit in `fading' phase for A4038]{A4038: {\it Left:} The best fit model spectrum in `fading' phase along with spectra in other phases. {\it Right:} The observed integrated spectrum of the relic with the best fit and the model spectra with $\Delta t_{4}\pm30\%$. See the electronic edition of the Journal for a color version 
of this figure.\label{fig:Fig. 14}}
\end{figure}

\begin{figure}
	\includegraphics[width=7.2 cm]{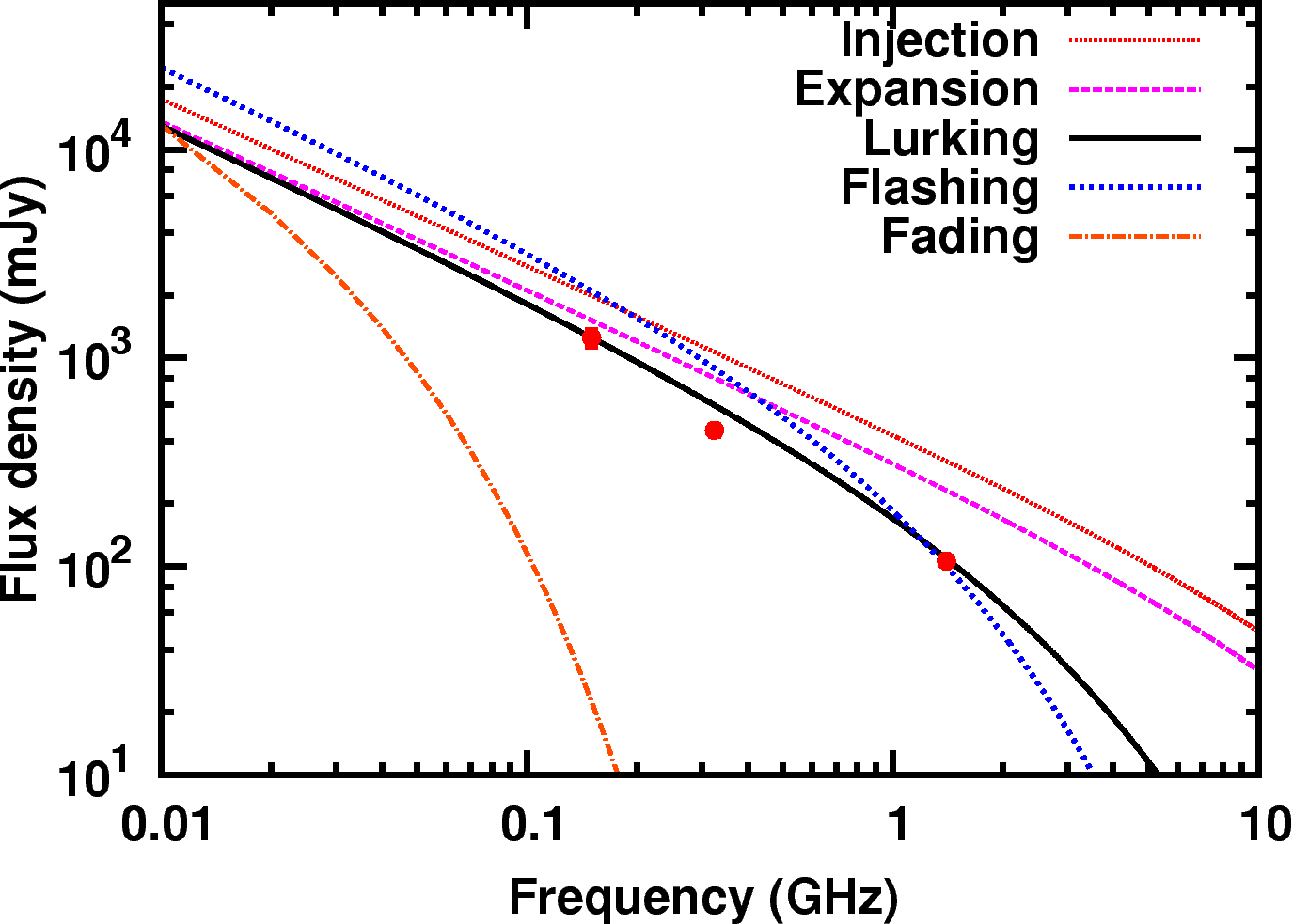}
	\includegraphics[width=7.2 cm]{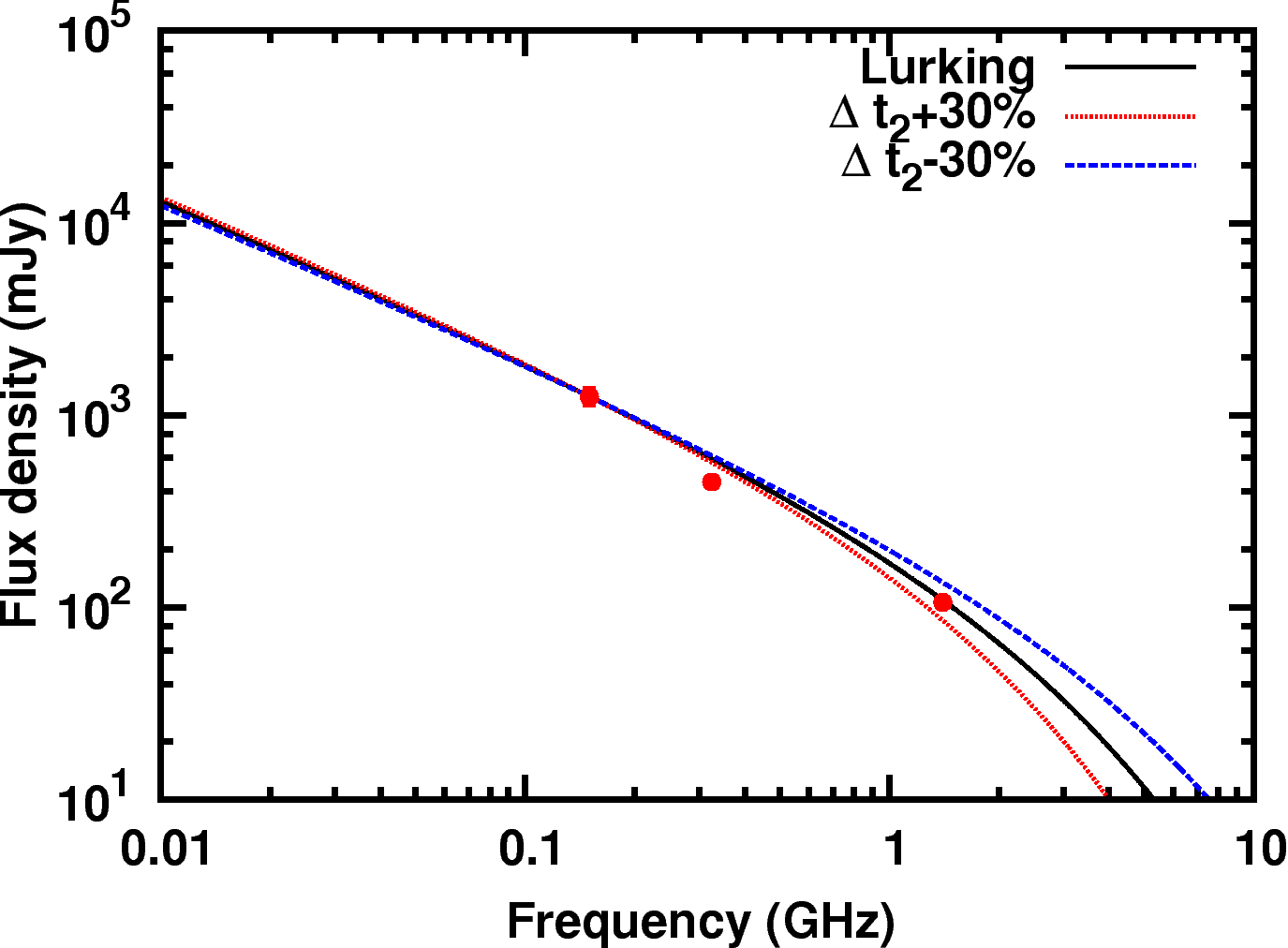}
\caption[Integrated spectrum and model Fits for A1664]{A1664: {\it Left:} The integrated 
spectrum of the relic in A1664 is plotted (Table 3). The lines are the spectra obtained from the EG01 model. Solid black line is the best fit (`lurking Phase') to the integrated spectrum of the relic in A664. Other lines are model spectra in the other phases of evolution. {\it Right:} The best fit model spectrum (black line) with the model spectra obtained when the duration of the best fit phase ($\Delta t_2$) was varied by $\pm30\%$. The integrated spectrum (points) is also plotted for reference. See the electronic edition of the Journal for a color version 
of this figure.}\label{fig:Fig. 15}
\end{figure}

\begin{figure}	
\includegraphics[width=7.2 cm]{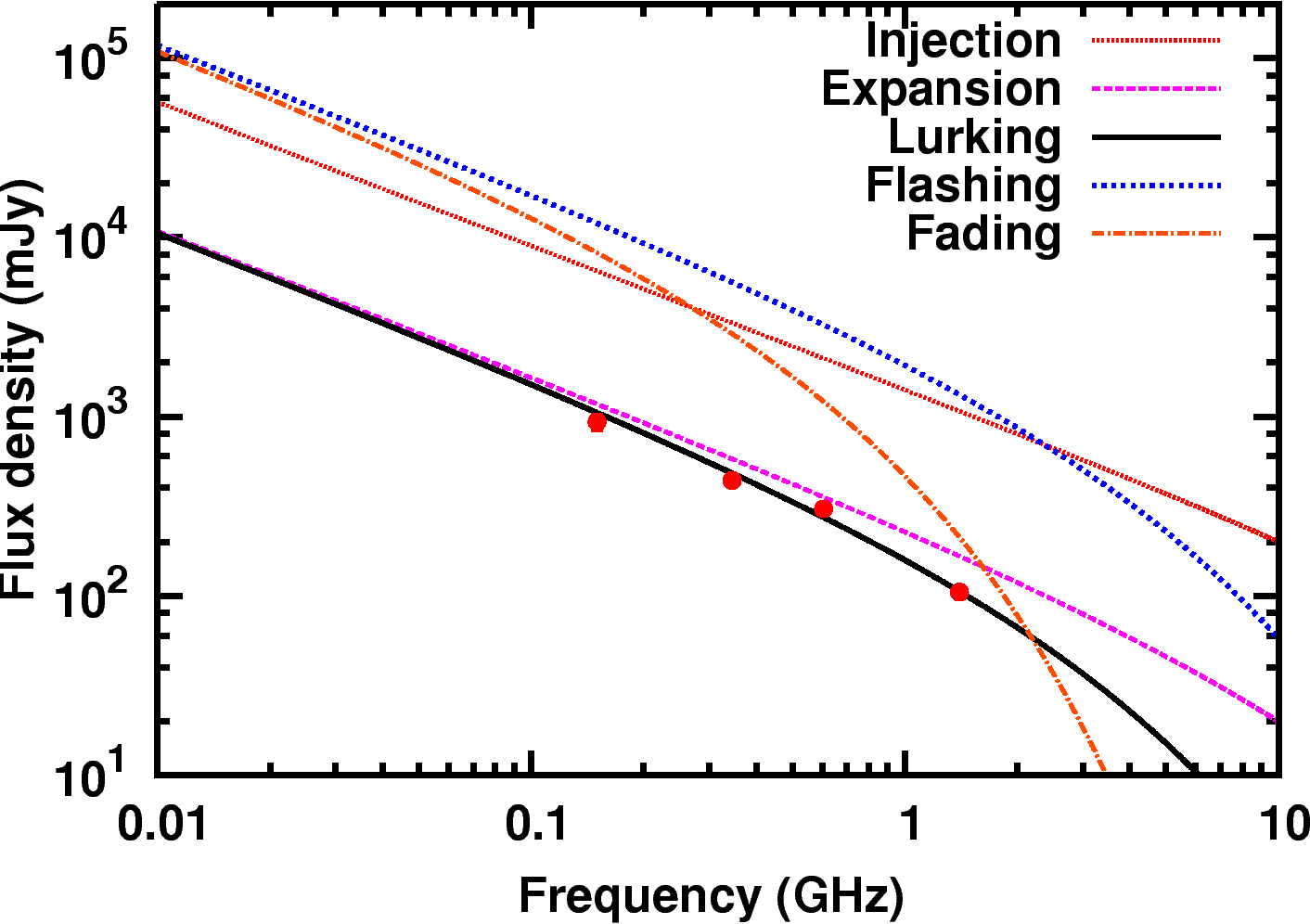}
\includegraphics[width=7.2 cm]{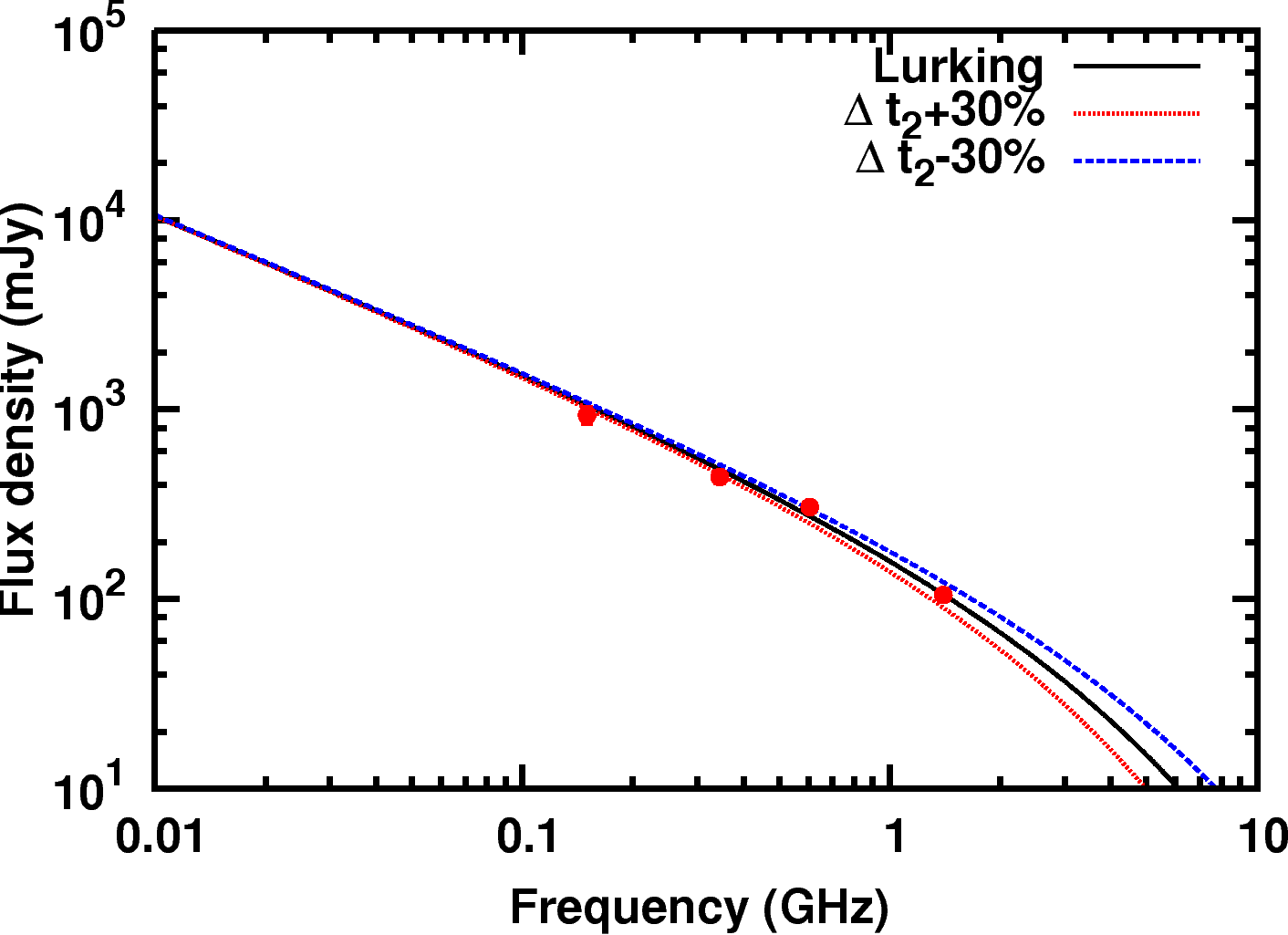}
\caption[Integrated spectrum of the relic near A786]{A786: Integrated spectrum of the relic
 near A786 is shown by points with error bars. {\it Left} Each curve is a spectrum in 
 a Phase (labelled in the legend) in the model. The solid black line is the best fit 
 spectrum in the `lurking' phase. {\it Right} The curves representing spectra when 
 the duration of the best fit phase was changed by $\pm30\%$. The best fit curve and the 
 integrated spectrum are plotted for reference. See the electronic edition of the Journal for a color version of this figure.\label{fig:Fig. 16}}
\end{figure}
\clearpage

\begin{figure}
	\includegraphics[width=8.0 cm]{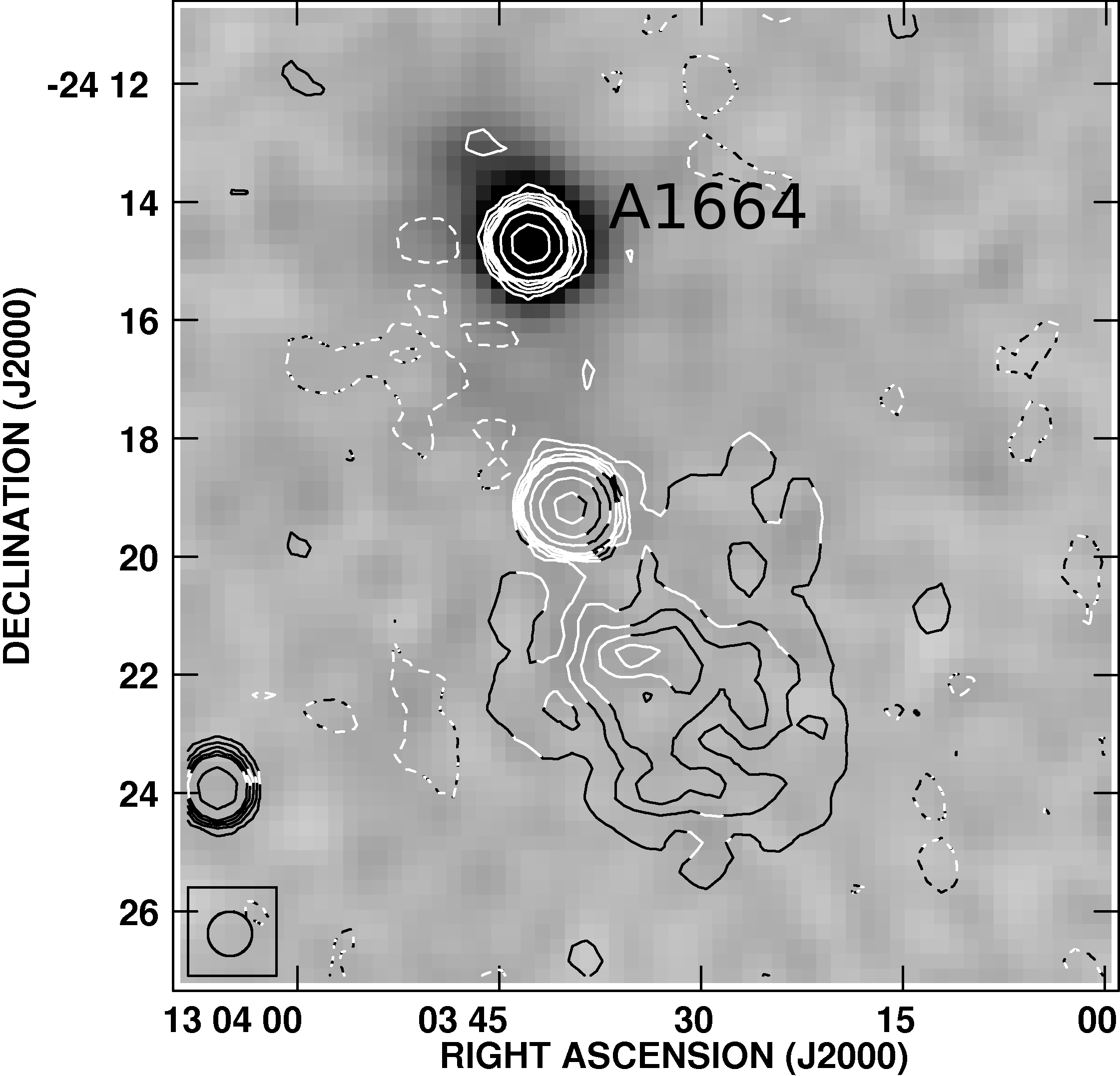}
\caption[Contours at 1400 MHz overlaid on ROSAT X-ray image of A1664]{A1664: 1400 MHz image 
(contours) overlaid on ROSAT (0.1-2.4 keV) image in grey-scale. Contour levels are at -1.05, 1.05, 2.10, 3.15, 4.20, 5.25, 10.50, 21.00, 42.00 \mjyb.\label{fig:Fig. 17}}
\end{figure}

\begin{figure}	
	\includegraphics[width=12.0 cm]{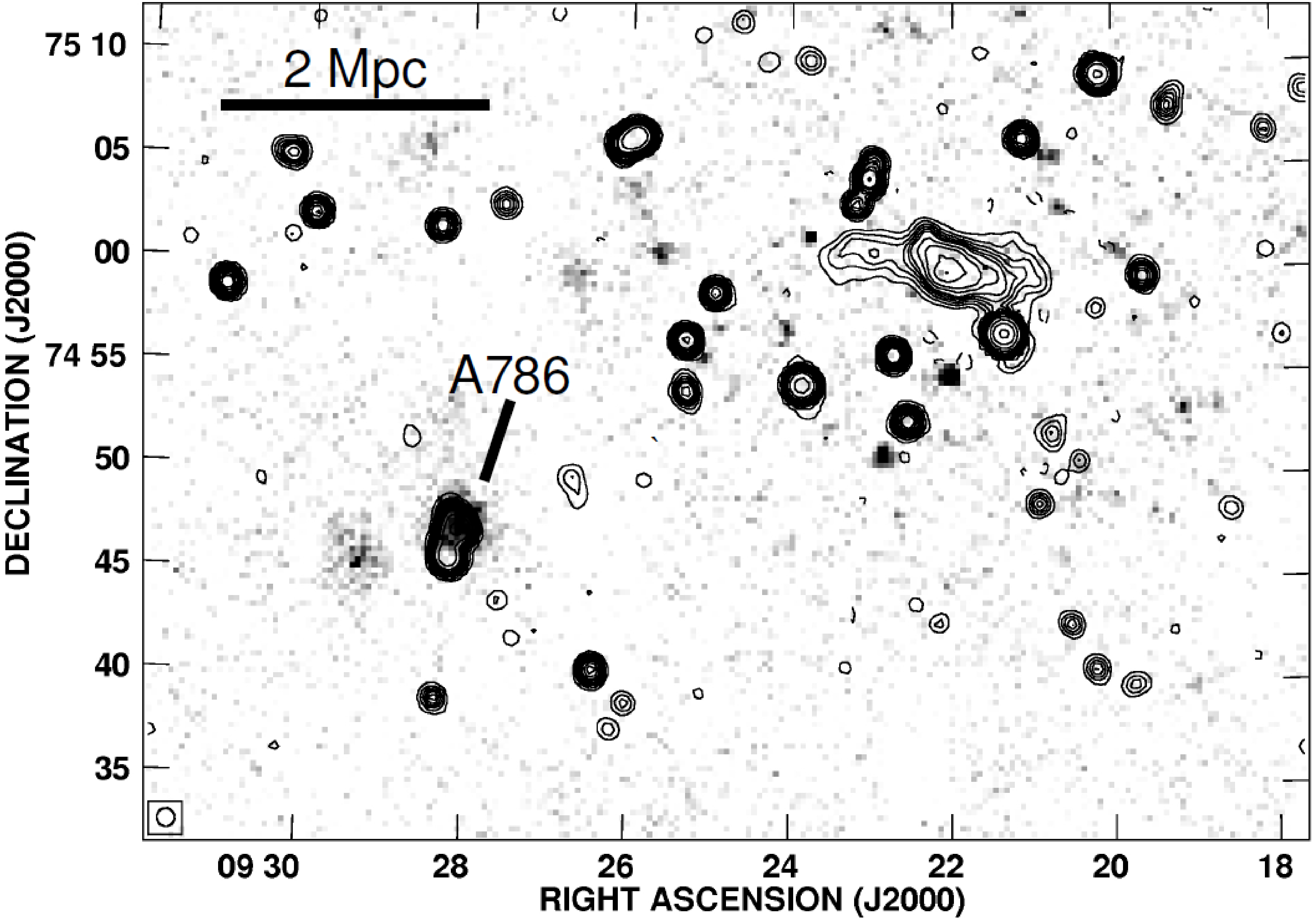}
	\caption[Image of the relic near A786 at 345 MHz overlaid on ROSAT]{A786: Contours at 345 MHz (WSRT) overlaid on the ROSAT HRI (0.1-2.4 keV) image. The cluster A786 lies to the southeast of the relic and is seen as a bright source in X-rays. A radio galaxy resides at the center of A786. Contour levels are at -2.4, 2.4, 4.8, 7.2, 9.6, 12.0, 14.4, 19.2, 24.0, 32.0, 40.0, 64.0, 128.0, 256.0, 512.0 \mjyb (synthesized beam$=52''\times51''$, P.A.$=7^\circ$)\label{fig:Fig. 18}}
\end{figure}


\begin{thebibliography}{}
\bibitem[Allen et al.(1995)]{all95}Allen, S. W., Fabian, A. C., Edge, A. C., Bohringer, H., \&  White, D. A. 1995, MNRAS, 275, 741
\bibitem[Bagchi et al.(2006)]{bag06}Bagchi, J., Durret, F., Neto, G. B. L. \& Paul, S.  2006, Science, 314, 791
\bibitem[Bonafede et al.(2009)]{bon09}Bonafede, A., Giovannini, G., Feretti, L., Govoni, F., \& Murgia, M. 2009, A\&A, 494, 429
\bibitem[Bridle et al.(1979)]{bri79}Bridle, A. H., Fomalont, E. B., Miley, G. K., \& Valentijn, E. A. 1979, \aap, 80, 201
\bibitem[Clarke \& Ensslin(2006)]{cla06}Clarke, T. E., \& Ensslin, T. A.  2006, AJ, 131, 2900
\bibitem[Cohen et al.(2007)]{coh07}Cohen, A. S., Lane, W. M., Cotton, W. D., Kassim, N. E., Lazio, T. J. W., Perley, R. A., Condon, J. J., \& Erickson, W. C. 2007, \aj, 134, 3, 1245
\bibitem[Cohen \& Clarke(2011)]{coh11}Cohen, A. S., \& Clarke, T. E. 2011, \aj, 141, 149
\bibitem[Condon et al.(1998)]{con98}Condon, J. J., Cotton, W. D., Greisen, E. W., 
    Yin, Q. F., Perley, R. A., Taylor, G. B., \& Broderick, J. J. 1998, \aj, 115, 1693
\bibitem[de Breuck et al.(2002)]{deb02}de Breuck, C. et al. 2002, \aj, 123, 637
\bibitem[Dewdney et al.(1991)]{dew91}Dewdney, P. E., Costain, C. H., McHardy, I., Willis, A. G., Harris, D. E. \& Stern, C. P. 
    1991, \apjs, 76, 1055
\bibitem[Dunn et al.(2005)]{dun05}Dunn, R. J. H., Fabian, A. C., \& Taylor, G. B. 2005, MNRAS, 364, 1343
\bibitem[Dunn \& Fabian(2006)]{dun06}Dunn, R. J. H. \& Fabian, A. C. 2006, \mnras, 373, 959
\bibitem[Ensslin et al.(1998)]{ens98}Ensslin, T. A., Biermann, P. L., Klein, U., \& Kohle, S. 1998, \aap, 332, 395
\bibitem[Ensslin \& Gopal-Krishna(2001)]{ens01}Ensslin, T. A., \& Gopal-Krishna 2001, A\&A, 366, 26 
\bibitem[Ensslin \& Bruggen(2002)]{ens02}Ensslin, T. A., \& Bruggen, M. 2002, \mnras, 331, 1011 
\bibitem[Giacintucci et al.(2008)]{gia08}Giacintucci, S. et al. 2008, A\&A, 486, 347
\bibitem[Giovannini et al.(1999)]{gio99}Giovannini, G., Tordi, M., \& Feretti, L. 1999, New Astr. Rev., 4, 141 
\bibitem[Govoni et al.(2001)]{gov01}Govoni, F., Feretti, L., Giovannini, G., Bohringer, H., Reiprich, T. H., \& Murgia, M. 2001, \aap, 376, 803
\bibitem[Harris et al.(1993)]{har93}Harris, D. E., Stern, C. P., Willis, A. G. \& Dewdney, P. E. 1993, \aj, 105, 769
\bibitem[Kale \& Dwarakanath(2009)]{kal09}Kale, R. \& Dwarakanath, K. S. 2009, \apj, 699, 1883
\bibitem[Kale \& Dwarakanath(2010)]{kal10}Kale, R. \& Dwarakanath, K. S. 2010, \apj, 718, 939
\bibitem[Kirkpatrick et al.(2009)]{kir09}Kirkpatrick, C. C. et al. 2009, \apj, 697, 867
\bibitem[Markevitch et al.(2005)]{mar05}Markevitch, M., Govoni, F., Brunetti, G., \& Jerius, D. 2005, \apj, 627, 733
\bibitem[McNamara(2002)]{mcn02}McNamara, B. R. 2002, The High Energy Universe at Sharp Focus: Chandra Science, ASP Conf. Proc., 262,  Ed. Schlegel, E. M. \& Vrtilek, S. D. San Francisco: Astronomical Society of the Pacific, 351
\bibitem[Orru et al.(2007)]{orr07}Orru, E., Murgia, M., Feretti, L., Govoni, F., Brunetti, G., Giovannini, G., Girardi, M., \& Setti, G.  2007, A\&A, 467, 943
\bibitem[Pierre \& Starck(1998)]{pie98}Pierre M., \& Starck J.-L. 1998, \aap, 330, 801
\bibitem[Pimbblet et al.(2006)]{pim06}Pimbblet, K. A., Smail, I., Edge, A. C., O'Hely, E., Couch, W. J., \& Zabludoff, A. 2006, \mnras, 366, 645
\bibitem[Randall et al.(2010)]{ran10}Randall, S. W., Clarke, T. E., Nulsen, P. E. J., Owers, M. S., Sarazin, C. L., Forman, W. R. , \& Murray, S. S. 2010, \apj, 722, 825
\bibitem[Reiprich \& Bohringer(2002)]{rei02}Reiprich, T. H., \& Bohringer, H. 2002, \apj, 567, 716
\bibitem[Slee \& Siegman(1983)]{sle83}Slee, O. B. \& Siegman, B. C. 1983,  1983, PASAu, 5, 114
\bibitem[Slee et al.(1994)]{sle94}Slee, O. B., Roy, A. L., \& Savage, A. 1994, AuJPh, 47, 145
\bibitem[Slee \& Reynolds(1984)]{sle84}Slee, O. B. \& Reynolds, J. E. 1984, PASAu, 5, 516
\bibitem[Slee \& Roy(1998)]{sle98}Slee, O. B., \& Roy, A. L. 1998, \mnras, 297L, 86
\bibitem[Slee et al.(2001)]{sle01}Slee, O. B., Roy, A. L., Murgia, M., Andernach, H., \& Ehle, M. 2001, \aj, 122, 1172
\bibitem[Struble and Rood(1999)]{str99}Struble, M. F., \& Rood, H. J. 1999, \apjs, 125, 35
\bibitem[van Weeren et al.(2009)]{wee09}van Weeren, R. J. et al.  2009, \aap, 506, 1083
\bibitem[van Weeren et al.(2010)]{wee10}van Weeren, R. J., Rottgering, H. J. A., Bruggen, M. \& Hoeft, M. 2010, Science, 330, 347
\bibitem[van Weeren et al.(2011a)]{wee11a}van Weeren, R. J., Hoeft, M., Rottgering, H. J. A., Bruggen, M., Intema, H. T., \& van Velzen, S. 2011, \aap, 528, 38
\bibitem[van Weeren et al.(2011b)]{wee11b}van Weeren, R. J., Rottgering, H. J. A., \& Bruggen, M. 2011, \aap, 527, 114
\end{thebibliography}
\end{document}